\documentclass[aps,prd,preprint,superscriptaddress,showpacs,preprintnumbers,amsmath,amssymb, unsortedadress]{revtex4-1}

 \usepackage{booktabs}
\usepackage{graphicx}% Include figure files
\usepackage{epstopdf}
\usepackage{dcolumn}% Align table columns on decimal point
\usepackage{float}
\usepackage[normalem]{ulem}
\usepackage{color}
\usepackage{dcolumn}% Align table columns on decimal point
 \usepackage{multirow}
\usepackage{subfigure}
\usepackage{cancel}
\newcommand{\be}{\begin{equation}}

\newcommand{\ee}{\end{equation}}
\newcommand{\bea}{\begin{eqnarray}}

\newcommand{\eea}{\end{eqnarray}}

\newcommand{\bfk}{\mbox{\boldmath $k$}}

\newcommand{\pup}{p^\uparrow}

\newcommand{\bfp}{\mbox{\boldmath $p$}}

\newcommand{\that}{\hat{t}} 
\newcommand{\uhat}{\hat{u}}
\newcommand{\shat}{\hat{s}}

\def\lsim{\mathrel{\rlap{\lower4pt\hbox{\hskip1pt$\sim$}}\raise1pt\hbox{$<$}}}
\def\gsim{\mathrel{\rlap{\lower4pt\hbox{\hskip1pt$\sim$}}\raise1pt\hbox{$>$}}}

\begin{document}

\title{Probing the Gluon Sivers Function through direct photon production at RHIC}

\author{Rohini M. Godbole}
\email{rohini@iisc.ac.in}
\affiliation{Centre for High Energy Physics, Indian Institute of Science, Bangalore, India.}
\affiliation{Institute of Physics and Astronomy, University of Amsterdam, The Netherlands}
\author{Abhiram Kaushik}
\email{abhiramk@iisc.ac.in}
\affiliation{Centre for High Energy Physics, Indian Institute of Science, Bangalore, India.}

\date{\today}

\author{Anuradha Misra}
\email{misra@physics.mu.ac.in}
\author{Siddhesh Padval}
\email{siddhesh.padval@physics.mu.ac.in}
\affiliation{Department of Physics, University of Mumbai, Mumbai, India.}

\begin{abstract}

We study the production of prompt-photons at RHIC in the context of a generalised parton model framework, with a view to obtain information on the gluon Sivers function (GSF). At RHIC energy ($\sqrt{s}=200$ GeV), the Compton process, $gq\to\gamma q$ contributes significantly to the production of direct-photons at midrapidity and dominates it in the negative (backward) rapdity region. We find that for direct photons, asymmetries of upto 10\% are allowed by a maximal gluon Sivers function. However, the asymmetry obtained using existing fits of the GSF available is literature is negligible.  We also estimate the impact that photons produced via fragmentation can have on the signal and find that their inclusion can dilute the asymmetry by between 10-50\% of the direct-photon value. Finally, using the Colour-Gauge Invariant generalised parton model (CGI-GPM) approach, we consider the effects of initial state and final state interactions which can affect the universality of the Sivers functions in different processes. We find that the inclusion of these effects leads to the size of the gluon contributions being roughly halved. However, in the backward region which we are interested in, the sizes of the quark contributions are suppressed even further, leading to increased dominance of the gluon contributions.

%Valid PACS numbers may be entered using the \verb+\pacs{#1}+ command.
\end{abstract}

%\keywords{Suggested keywords}%
\maketitle

\section{\label{intro}Introduction}
Transverse single-spin asymmmetries (SSA) can provide information on the three-dimensional structure of hadrons. They have hence been a subject of great interest in recent times. In the past few years, a large amount of data on SSAs have become available in a wide variety of processes such as semi-inclusive deep-inelastic scattering, hadroproduction of light and heavy mesons (see Refs.~\cite{DAlesio:2007bjf, Barone:2010zz} for reviews of experimental data on the subject)  and most recently in Drell-Yan~\cite{Franco:2018wtp}. One of the theoretical approaches used to describe these asymmetries is TMD factorisation~\cite{Ji:2004xq,Ji:2004wu,Bacchetta:2008xw,Collins:2011zzd}. In this approach, factorisation in terms of transverse momentum dependent parton distribution functions (TMD-PDF) and fragmentation functions (TMD-FF) is assumed. These functions depend on the transverse-momentum of the parton in addition to the light-cone momentum fraction, i.e they are of the form $f_{i/h}(x,\bfk,Q)$ and $D_{h/i}(z,\bfk,Q)$ respectively. This is in contrast to the commonly used collinear PDFs and FFs, which depend only on the light-cone momentum fraction as the transverse-momentum of the parton is integrated over. So far, TMD factorisation has been demonstrated only for processes which have two scales --- a hard, high energy scale such as the virtuality of the photon in the Drell-Yan process and a relatively soft scale of the order of $\Lambda_\text{QCD}$, such as the transverse momentum of the Drell-Yan lepton-pair. In the TMD approach, one of the main TMDs that can lead to an SSA is the Sivers distribution~\cite{Sivers:1989cc, Sivers:1990fh}. This encodes the correlation between the azimuthal anisotropy in the distribution of an unpolarised parton and the spin of its parent hadron. This anisotropy in the parton's transverse momentum distribution can lead to an azimuthal anisotropy in the distribution of the inclusive final state, i.e., a SSA. 

Though TMD factorisation has not been formally established for hard single-scale processes such as $p^\uparrow p \to h+X$ and $p^\uparrow p\to\gamma+X$, an effective description of SSAs in such processes in terms of the TMDs --- under the assumption of factorisation and universality --- has been phenomenologically succesful~\cite{Anselmino:2011ch, Anselmino:2012rq,Anselmino:2013rya,Anselmino:1994tv,DAlesio:2004eso,Anselmino:2005sh}. This effective description is commonly referred to in literature as the Generalised Parton Model (GPM).

Recently a modification of the Generalised Parton Model has been proposed in which the process dependence of the Sivers function is taken into account. In this approach, known as the Colour-Gauge Invariant Generalised Parton Model (CGI-GPM)~\cite{Gamberg:2010tj, DAlesio:2011kkm, DAlesio:2013cfy}, the process dependent initial state interactions (ISIs) and final state interactions (FSIs) are treated using one-gluon exchange approximation. These interactions provide the complex phase necessary for the SSA. The process dependence of the Sivers functions, that arises from these interactions, is then shifted onto appropriately defined `modified' partonic cross-sections and the Sivers functions can still be treated as universal. This approach was first proposed in Ref.~\cite{Gamberg:2010tj}, where they considered quark Sivers functions, and has recently been extended to include gluon Sivers functions in Ref.~\cite{DAlesio:2017rzj}. The process dependence of the Sivers function (and in general T-odd functions) was earlier studied in the context of two-scale scale processes in Refs.~\cite{Bacchetta:2005rm, Bomhof:2006dp}.

While the quark Sivers functions have been widely studied over the years, the gluon Sivers function (GSF) still remains poorly measured. An indirect estimate of the gluon Sivers function was obtained using a GPM framework in Ref.~\cite{DAlesio:2015fwo} where they fit the gluon Sivers function to midrapidity data on SSA in $\pi^0$ production at RHIC. In the analysis, the quark contribution to the SSA was calculated using QSFs as extracted from semi-inclusive deep inelastic scattering data. The {\it small and positive} GSF fits obtained by the analysis predicted asymmetries much smaller than allowed by the positivity bound on the GSF. The said bound restricts the GSF to be less than twice the unpolarised TMD gluon distribution. Further, a recent study of large-$p_T$ hadron pair production in COMPASS indicates a \textit{substantial, negative} gluon Sivers asymmetry but with large errors and hence consistent with zero at 2.5 sigma level for a proton target~\cite{Adolph:2017pgv}. Large-$p_T$ hardon pairs are produced in this process through photon-gluon fusion, a process which gives direct access to the gluon content of the proton. The differences between these two results, though of limited statistical significance, make it clear that the GSF needs to be studied in more detail and with unambiguous probes.

%Both of these probes suffer from theoretical uncertainties such as the validity of factorisation, and also the impact of initial and final state effects, which can break the universality of the GSF.  Overall the situation regarding the GSF is really quite unclear.}

More direct probes of the GSF are thus needed. { Closed and open}  heavy-flavour production offer such probes. A GPM study of open charm production as a probe of the GSF was proposed in Ref.~\cite{Anselmino:2004nk} for the process $p^\uparrow p\to D^0+X$. Therein they considered two extreme scenarios for the GSF: zero and saturated. By `saturated' we mean the Sivers function with its positivity bound of twice the unpolarised TMD, i.e., $|\Delta^Nf_{i/p^\uparrow}(x,\mathbf{k}_\perp)|/2f_{i/p}(x,\mathbf{k}_\perp)\leq1$, saturated for all values of $x$.  Their study indicated that an observation of SSA for this process at RHIC,  can give a direct indication of a nonzero gluon Sivers function. Further in Ref.~\cite{Godbole:2016tvq} we calculated the SSA for the same process (open charm hadroproduction) using the fits of Ref.~\cite{DAlesio:2015fwo} and found that these fits predict sizeable, measurable asymmetries. In Ref.~\cite{Godbole:2017fab}, we proposed the low-virtuality leptoproduction of open charm and studied it in the context of the GPM framework. Unlike hadroproduction, this process does not involve  any  contributions from quarks and hence can probe the gluon Sivers function effectively. In this case too we found  results similar to those  for the hadroproduction of open charm. For the kinematics of COMPASS and a future Electron-Ion Collider (EIC), the fits of Ref.~\cite{DAlesio:2015fwo} gave sizable and distinct asymmetries. Further we found that the asymmetry was well-preserved in the kinematics of the muons decaying from the $D$-meson.

Apart from the production of open charm, probes involving the production of closed charm, i.e. $J/\psi$, can also give direct access to the gluon content of the proton. The low-virtuality leptoproduction of closed charm, i.e. $J/\psi$ was  proposed as a probe of the GSF in Refs.~\cite{Godbole:2012bx, Godbole:2013bca,Godbole:2014tha} and the hadroproduction of closed charm was studied in Refs.~\cite{Godbole:2017syo, DAlesio:2017rzj}. Recently, the PHENIX collaboration at RHIC has measured the SSA in the production of $J/\psi$ in $p^\uparrow p$ collisions~\cite{Aidala:2018gmp}. They find that the data indicate a positive asymmetry at the two standard-deviation level in the $x_F<0$ region.

In this work, using both the GPM and CGI-GPM approaches, we study the hadroproduction of prompt-photons as a possible probe of the gluon Sivers function. By prompt-photons we mean both,  the direct photons which are created in the hard process and fragmentation photons which are created by fragmentation of outgoing partons from the hard-process. At LO, direct photons are produced through the fundamental 2-to-2 hard scattering subprocesses, $gq\to\gamma q$ and $q\bar q\to \gamma+g$. The first of these subprocesses, the QCD Compton process, dominates in $pp$ collisions. Indeed direct-photon data from fixed target experiments were used in early global fits of the  collinear PDFs in unpolarised protons  to constrain the gluon component~\cite{Morfin:1990ck, Harriman:1990hi, Diemoz:1987xu}. At RHIC, study of prompt photon production in the midrapidity and backwards rapidity regions should give clean and direct access to the gluon content of the polarised proton, due to the dominance of the QCD Compton process. Since the photon is produced in  the hard process and is colourless, the probe is unaffected by theoretical uncertainties related to hadronization or final state interactions. For this reason, SSA in the production of direct-photons in the backward region was first proposed as a probe of the gluon Sivers function by Schmidt, Soffer and Yang~\cite{Schmidt:2005gv} (SSY). They suggested in general the large-$P_T$ region in the backward hemisphere. The SSA in this process has also been studied in the context of an alternative mechanism in the colour glass condensate formalism and found to be zero~\cite{Kovchegov:2012ga}.

In this work, we follow up on the work of SSY and consider estimates of the asymmetry in the production of direct photons in the backwards, i.e., negative rapidity region. In their work, SSY neglected the partonic transverse momenta in the hard-part. However it is known that partonic transverse momenta cannot be neglected in the hard-part. They lead to an important consequence, {\bf \it viz.}  a suppression of the SSA in the backward region relative to the forward region, as was found for the case of open-charm production in Ref.~\cite{Anselmino:2004nk}. Here, we include the effect of partonic transverse momenta in the hard process. We find that after the inclusion of these effects, the dominance of the gluon contribution in the backward region continues even though the SSAs are suppressed. In this work, following our earlier studies of the GSF in open charm production~\cite{Godbole:2016tvq, Godbole:2017fab}, we consider estimates for the asymmetry obtained using saturated quark and gluon Sivers functions. The use of saturated QSFs and GSF gives the upper bound on the possible asymmetry and further allows us to study the general kinematic dependencies of the asymmetry and the relative importances of the quark and gluon contributions to the asymmetry. It also allows us to assess the sensitivity of the probe to the uncertainties in our current knowledge of the collinear PDFs. Further we also consider the contribution to the asymmetry from photons produced via the fragmentation of partons. While the  contribution to the signal and hence the asymmetry coming from the fragmentation component can be reduced to an extent by applying a photon isolation requirement, it cannot be completely eliminated. Hence it is important to study the impact that it can have on the asymmetry and hence  on this probe of the GSF. Therefore, we also consider the asymmetry in the inclusive (direct as well as fragmentation) photons and study how it differs from the direct photon asymmetry. 

We then consider existing fits of the QSFs and the GSF to  give predictions for the expected asymmetry. We consider two fits of the QSFs~\cite{Anselmino:2005ea, Anselmino:2008sga}, both  of which have been obtained by fitting to  data on semi-inclusive deep-inelastic scattering. Associated with these two QSF sets are two fits of the GSF in Ref.~\cite{DAlesio:2015fwo} which, as mentioned earlier, were obtained by fitting to the  data on SSA in midrapidity $\pi^0$ production at RHIC. Further we also consider indirect bounds on the gluon Sivers function based on the Burkardt Sum Rule (BSR)~\cite{Burkardt:2004ur}. The BSR is the requirement that the net transverse momenta of all partons in a transversely polarised proton must vanish. The latest fits of the QSFs from Ref.~\cite{Anselmino:2008sga} allow a gluon transverse momentum in the range $-10\leq\langle k_{\perp g}\rangle\leq48\text{ MeV}$. This constraint on the allowed transverse momentum can be used to constrain the size of the gluon contribution to the asymmetry. We plot the asymmetry values allowed by the BSR constraint, along with the predictions from the fits.

Finally we study how the results of the above analysis are affected when we take into account the effects of initial and final state interactions using the CGI-GPM approach. We do so for both the direct as well as { the}  fragmentation contribution to the asymmetry.

This paper is organised as follows: In section II, we give expressions for the relevant quantities in the GPM framework. In section III, we present the CGI-GPM formalism and give the modified hard-part for the relevant processes. In section IV we present the parametrisation of  the various TMDs in the analysis as well as details of the  QSF and the GSF parameters used.  Finally, in section V  we present estimates of the asymmetry in both the GPM and CGI-GPM frameworks.

\section{Prompt photon production in The GPM Formalism}

Prompt photons can be produced either in the hard scattering, or through the fragmentation of a final state parton into a photon. We refer to the former as {\it direct} photons and the latter as {\it fragmentation} photons. At leading order $\mathcal{O}(\alpha_s\alpha_\text{em})$, direct photons are produced through the QCD Compton process, $gq\to\gamma q$ and quark-antiquark annihilation into a photon and a gluon, $q\bar q \to \gamma g$. Of these $gq\to\gamma q$ dominates at RHIC energy and hence the production of direct photons can give direct access to the gluon content of the proton.

Fragmentation photons can be produced at leading order, $\mathcal{O}(\alpha_s^2\alpha_\text{em})$ through the standard 2-to-2 QCD parton scattering processes with the final state parton fragmenting into a photon. Here, unlike direct photon production, the partonic subprocess is at $\mathcal{O}(\alpha_s^2)$ instead of $\mathcal{O}(\alpha_s\alpha_\text{em})$. Though this is naively higher order in $\alpha_s$ as compared to direct photon production, the parton-to-photon fragmentation functions (FFs) grow logarithmically with $Q^2$, making them effectively of order $1/\alpha_s$. This logarithmic growth of the parton-to-photon FFs makes the production of fragmentation photons effectively at the same leading order $\mathcal{O}(\alpha_s\alpha_\text{em})$ as the production  of direct photons. Though the fragmentation photon contribution can be discriminated by applying a photon isolation requirement, it can never be completely eliminated. Therefore it is important to consider the effect of fragmentation photons on the observed asymmetry in the consideration of SSAs in an inclusive signal. An illustration of both mechanisms of prompt-photon production is given in Fig.~1.

\begin{figure}[hbt] 

	\centering
	
	\includegraphics[width=0.95\linewidth]{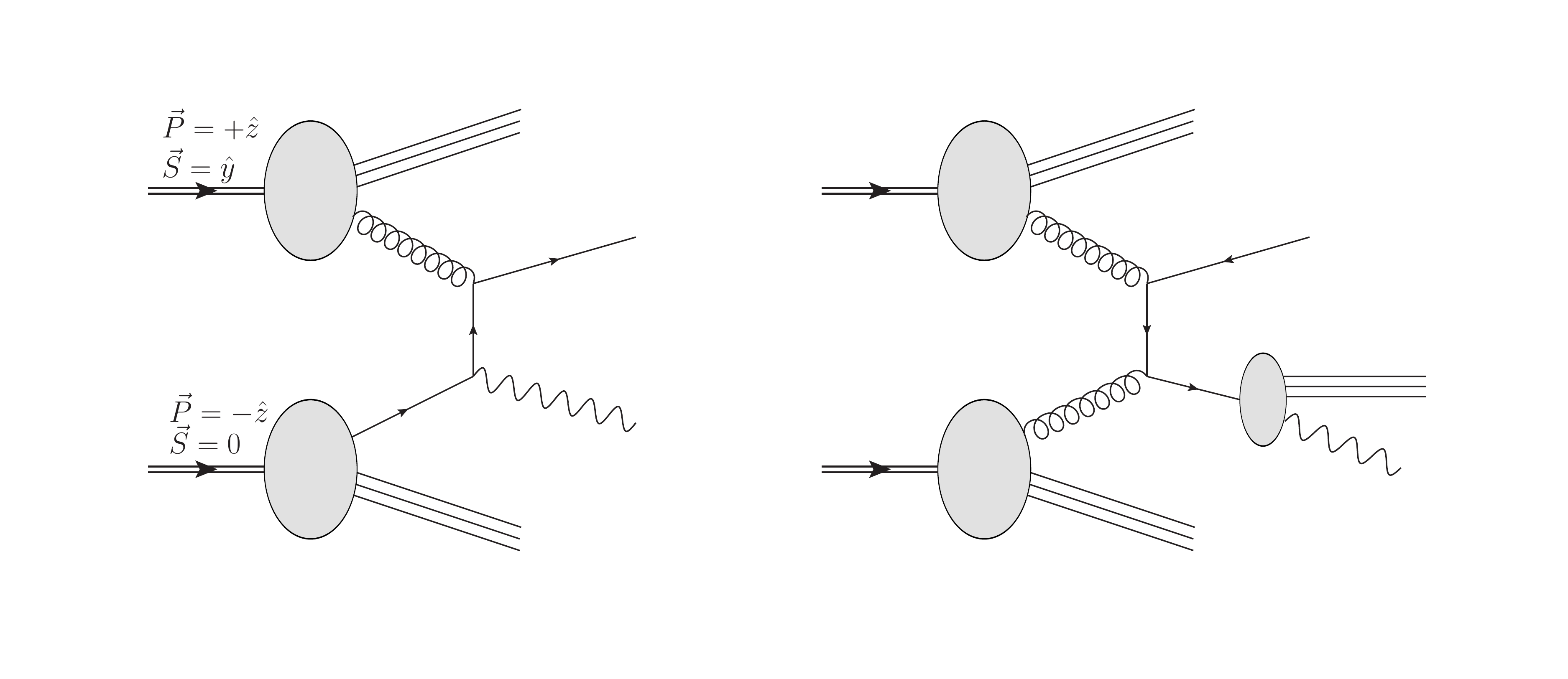}
	\vspace*{-1cm}
	\caption{Representative diagrams for prompt photon production at the hard scattering (left) and in the fragmentation of a final state parton (right), in hadron-hadron collisions. We consider one of the hadrons to be a proton moving in the +Z direction, with a polarisation along the +Y axis. The other hadron is an unpolarised proton moving along the -Z direction.}
	\label{fig:directandfrag}
\end{figure}

In this work, we are concerned with the single-spin asymmetry in the hadroproduction of prompt-photons, 
\be
A_N=\frac{d\sigma^\uparrow-d\sigma^\downarrow}{d\sigma^\uparrow+d\sigma^\downarrow}
\label{SSA}
\ee
where $d\sigma^{\uparrow( \downarrow)}$ is the invariant differential cross-section for the process $p^{\uparrow(\downarrow)} p\to \gamma+X$ with the spin of the transversely polarised proton being aligned in the $\uparrow$($\downarrow$) direction with respect to the production plane. Here, $\uparrow$ would be the $+Y$ direction in a frame where the polarised proton is moving along the $+Z$ direction and the photon is produced in the $XZ$ plane.

In the following, we give the expressions  for  the denominator and numerator of Eq.~\ref{SSA} for the case of both direct photons as well as fragmentation photons.

\subsection{Direct photon production}
For {\it direct photons}, we can write the denominator and numerator of Eq.~1 as,
\bea
d\sigma ^\uparrow &+& d\sigma ^\downarrow = \frac{E_\gamma \, d\sigma^{p^\uparrow p \to \gamma X}} {d^{3} \bfp_\gamma} +
\frac{E_\gamma \, d\sigma^{p^\downarrow p \to \gamma X}} {d^{3} \bfp_\gamma} \\ \nonumber
&=&2\sum_{a,b=g,q,\bar q}\int dx_a d^2\bfk_{\perp a}dx_b d^2\bfk_{\perp b}f_{a/p}(x_a, k_{\perp a}) f_{b/p}(x_b,k_{\perp b})\frac{\hat s}{x_ax_bs}\frac{d\hat\sigma^{ab\to\gamma d}}{d\that}\frac{\hat s}{\pi}\delta(\hat s+\hat t+\hat u)
\label{directden}
\eea
and
\bea
d\sigma ^\uparrow &-& d\sigma ^\downarrow = \frac{E_\gamma \, d\sigma^{p^\uparrow p \to \gamma X}} {d^{3} \bfp_\gamma} -
\frac{E_\gamma \, d\sigma^{p^\downarrow p \to \gamma X}} {d^{3} \bfp_\gamma} \\ \nonumber
&=&\sum_{a,b=g,q,\bar q}\int dx_a d^2\bfk_{\perp a}dx_b d^2\bfk_{\perp b}~\Delta^Nf_{a/p^\uparrow}(x_a, \bfk_{\perp a}) f_{b/p}(x_b,k_{\perp b})\frac{\hat s}{x_ax_bs}\frac{d\hat\sigma^{ab\to\gamma d}}{d\that}\frac{\hat s}{\pi}\delta(\hat s+\hat t+\hat u)
\label{directnum}
\eea
In the above expressions, $x_a$ and $x_b$  are the light-cone momentum fractions of the incoming partons of  the polarised and unpolarised proton respectively.  $\bfk_{\perp a}$ and $\bfk_{\perp b}$ are the transverse momenta of the partons $a$ and $b$. $\shat=(p_a+p_b)^2$, $\that=(p_\gamma-p_a)^2$ and $\uhat=(p_\gamma-p_b)^2$ are the Mandelstam variables for the relevant subprocesses: {the QCD} Compton scattering process $gq\to\gamma q$, {as well as}  $q\bar q\to \gamma g$.

The expressions $\Delta^Nf_{i/p^\uparrow}(x,\mathbf{k}_\perp)$ and $f_{i/p}(x,\mathbf{k}_\perp)$ are the Sivers function and unpolarised TMD for parton $i$, respectively. The functional forms used for these two distributions are given in Sec. IV.

The Sivers function, $\Delta^Nf_{i/p^\uparrow}(x,k_\perp;Q)$ describes the azimuthal anisotropy in the transverse momentum distribution of an unpolarised parton, in a transversely polarised hadron,
\bea
f_{i/h^\uparrow}(x,\mathbf{k}_\perp,\mathbf{S};Q)&=&f_{i/h}(x,k_\perp;Q)+\frac{1}{2}\Delta^N f_{i/h^\uparrow}(x,k_\perp;Q)\frac{\epsilon_{ab}k_\perp^a S^b}{k_\perp}\nonumber \\
&=&f_{i/h}(x,k_\perp;Q)+\frac{1}{2}\Delta^N f_{i/h^\uparrow}(x,k_\perp;Q)\cos\phi_{\perp}
\label{Sivers}
\eea
where $\mathbf{k}_\perp=k_\perp(\cos\phi_\perp, \sin\phi_\perp)$. Another notation, $f_{1T}^{\perp i}$, is also commonly used for the Sivers function and is related to $\Delta^Nf_{i/h^\uparrow}$ by,
\be
\Delta^Nf_{i/h^\uparrow}(x,k_\perp)=-2\frac{k_\perp}{M_h}f^{\perp i}_{1T}(x,k_\perp).
\ee
where $M_h$ is the mass of the hadron, which in this case, is the proton. We will use this notation when discussing the CGI-GPM formalism in Sec.~III.

The partonic cross-sections can be written as
\be
\frac{d\hat\sigma^{ab\to\gamma d}}{d\that}=\frac{\pi\alpha_s\alpha_\text{em}}{\shat^2}H^U_{ab\to\gamma d},
\ee
with the hard-parts for the two subprocesses given by,
\be
H^U_{gq\to\gamma q}=-\frac{e_q^2}{3}\left[\frac{\uhat}{\shat}+\frac{\shat}{\uhat}\right], \hspace*{1cm}H^U_{q\bar q\to\gamma g}=\frac{8}{9}~e_q^2\left[\frac{\uhat}{\that}+\frac{\that}{\uhat}\right]
\ee

The on-shell condition $\shat+\that+\uhat=0$, can be used to fix one of the integration variables, in this case, $x_b$.

\subsection{Photons from fragmentation of quarks}
For photons that are produced in the fragmentation of quarks, there can be two TMDs that contribute towards a SSA, the Sivers function and the Collins function~\cite{Collins:1992kk}. Here we consider only the Sivers function. A model calculation of the quark-to-photon Collins function has shown that the contribution to the asymmetry from the Collins function is negligible~\cite{Gamberg:2012iq}. Following the formalism used in single-inclusive hadron productions~\cite{DAlesio:2004eso,Godbole:2016tvq}, the denominator and numerator of Eq.~1 can be written as,
\bea
d\sigma ^\uparrow &+& d\sigma ^\downarrow = \frac{E_\gamma \, d\sigma^{p^\uparrow p \to \gamma X}} {d^{3} \bfp_\gamma} +
\frac{E_\gamma \, d\sigma^{p^\downarrow p \to \gamma X}} {d^{3} \bfp_\gamma} \\ \nonumber
&=&2\sum_{a,b=g,q,\bar q}\int dx_a d^2\bfk_{\perp a}dx_b d^2\bfk_{\perp b}dzd^3\bfk_\gamma\delta(\bfk_\gamma.\hat\bfp_q)f_{a/p}(x_a, k_{\perp a}) f_{b/p}(x_b,k_{\perp b})\\ \nonumber
&\times& \frac{\hat s}{x_ax_bs}\frac{d\hat\sigma^{ab\to c d}}{d\that}\frac{\hat s}{\pi}\delta(\hat s+\hat t+\hat u)\frac{1}{z^2}J(z,|\bfk_\gamma|)D_{\gamma/q}(z,\bfk_\gamma)
\label{fragden}
\eea
and
\bea
d\sigma ^\uparrow &-& d\sigma ^\downarrow = \frac{E_\gamma \, d\sigma^{p^\uparrow p \to \gamma X}} {d^{3} \bfp_\gamma} -
\frac{E_\gamma \, d\sigma^{p^\downarrow p \to \gamma X}} {d^{3} \bfp_\gamma} \\ \nonumber
&=&\sum_{a,b=g,q,\bar q}\int dx_a d^2\bfk_{\perp a}dx_b d^2\bfk_{\perp b}dzd^3\bfk_\gamma\delta(\bfk_\gamma.\hat\bfp_q)\Delta^Nf_{a/p^\uparrow
}(x_a, k_{\perp a}) f_{b/p}(x_b,k_{\perp b})\\ \nonumber
&\times& \frac{\hat s}{x_ax_bs}\frac{d\hat\sigma^{ab\to c d}}{d\that}\frac{\hat s}{\pi}\delta(\hat s+\hat t+\hat u)\frac{1}{z^2}J(z,|\bfk_\gamma|)D_{\gamma/q}(z,\bfk_\gamma).
\label{fragnum}
\eea

In the above expressions $D_{\gamma/q}(z,\bfk_\gamma)$ is the TMD fragmentation function decribing the fragmentation of the quark $q$ into a photon carrying a light-cone momentum fraction $z=p_\gamma^+/p_q^+$,  and  a transverse momentum $\bfk_\gamma$ with respect to the fragmenting quark direction. 
$J(z,|\bfk_\gamma|)$ is the Jacobian factor connecting the phase-space of parton $c$ to the phase-space of the photon. It is given by,
\be
J(z,|\bfk_\gamma|)=\frac{(E_\gamma+\sqrt{E_\gamma^2-k_\gamma^2})^2}{4(E_\gamma^2-k_\gamma^2)}.
\ee
The delta function $\delta (\bfk_\gamma \cdot \hat{\bfp}_q)$ in Eqs.~\ref{fragden} and \ref{fragnum} confines the integration region for $\bfk_\gamma$ to the two-dimensional plane perpendicular to the direction of the fragmenting quark $\hat\bfp_c$, i.e., 
\be
\int d^3\bfk_\gamma~\delta (\bfk_\gamma \cdot \hat{\bfp}_c)D_{\gamma/c}(z,\bfk_\gamma)...=\int d^2\bfk_{\perp\gamma}D_{D/c}(z,\bfk_{\perp\gamma})...
\ee
where $\bfk_{\perp \gamma}$ represents values of transverse momenta on the allowed plane.
For photon production via fragmentation, at LO the partonic cross-section is of order $\alpha_s^2$,
\be
\frac{d\hat\sigma^{ab\to c d}}{d\that}=\frac{\pi\alpha_s^2}{\shat^2}H^U_{ab\to c d},
\ee
and the relevant hard-parts can be found, for instance in Ref.~\cite{Owens:1986mp}.

The details of the treatment of parton kinematics for both direct and fragmentation processes are presented in the appendix.

\section{The CGI-GPM formalism}
In the generalised parton model it is assumed that all the transverse-momentum-dependent densities are universal. For instance, one can use quark Sivers function fitted to SIDIS data, to calculate asymmetries in hadroproduction processes~\cite{DAlesio:2015fwo}.  Similarly the GSF fitted to data on pion-hadroproduction can be used to calculate the asymmetry in $J/\psi$ production~\cite{Godbole:2017syo}.  However, different processes --- take for example, SIDIS and Drell-Yan --- can have different initial and final state interactions between the active partons and the spectators from the polarised proton. For instance, in SIDIS, the scattered quark can exchange soft gluons with the remnant of the proton. This would be a final state interaction (FSI). In case of Drell-Yan, the incoming quark from the unpolarised proton can exchange soft gluons with the transversely polarised proton. This would be an initial state interaction (ISI). These interactions can affect the universality of the TMD densities. The effects of these different ISI/FSIs can be understood by looking at the different Wilson line configurations that are required to render the operator definition of the Sivers function gauge invariant. 
%\textcolor{blue}{Maybe put in here examples from SIDIS and DY.}

\begin{figure}
%	\centering
% \begin{adjustwidth}{-\oddsidemargin-1in}{-\rightmargin}
%\hspace*{-1.5cm}
 \advance\leftskip-0.5cm
\minipage{0.25\paperwidth}
%	\begin{subfigure}[b]{0.5\paperwidth}
	\includegraphics[width=0.8\linewidth, trim={5cm 16cm 5cm 5cm}]{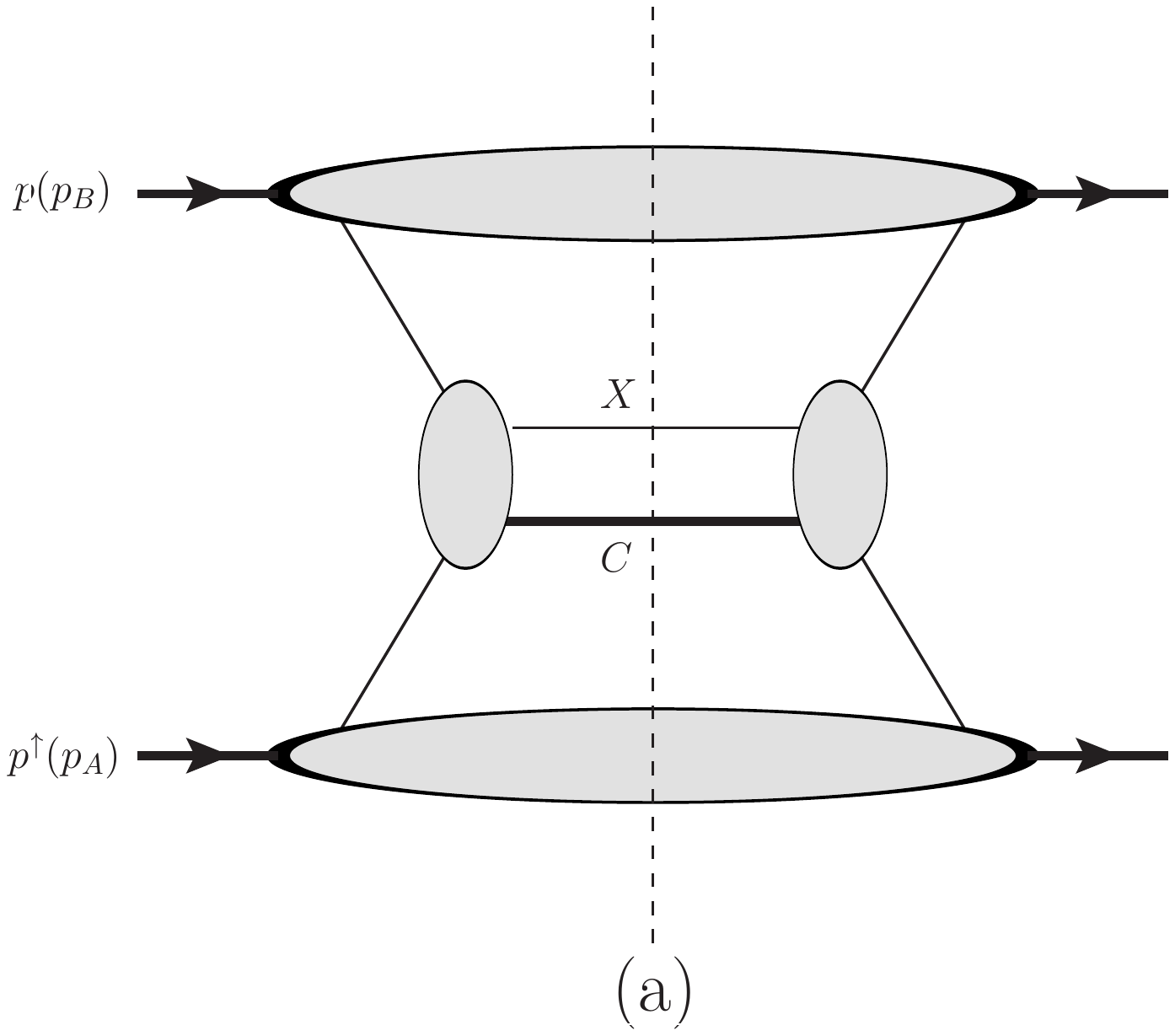}
%	\end{subfigure}
\endminipage\hfill
\minipage{0.25\paperwidth}
	\includegraphics[width=0.8\linewidth, trim={5cm 16cm 5cm 5cm}]{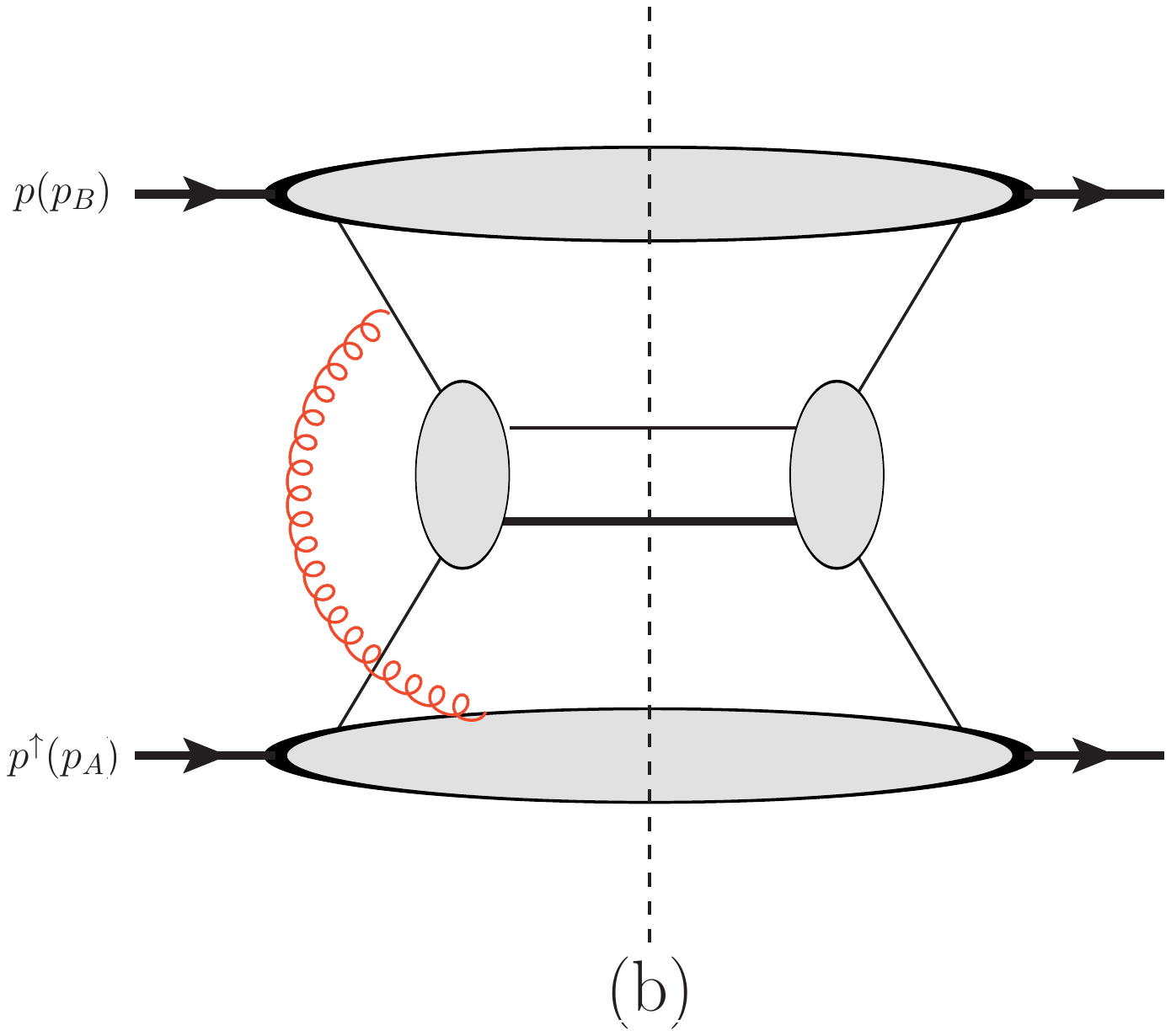}
\endminipage\hfill
\minipage{0.25\paperwidth}
	\includegraphics[width=0.8\linewidth, trim={5cm 16cm 5cm 5cm}]{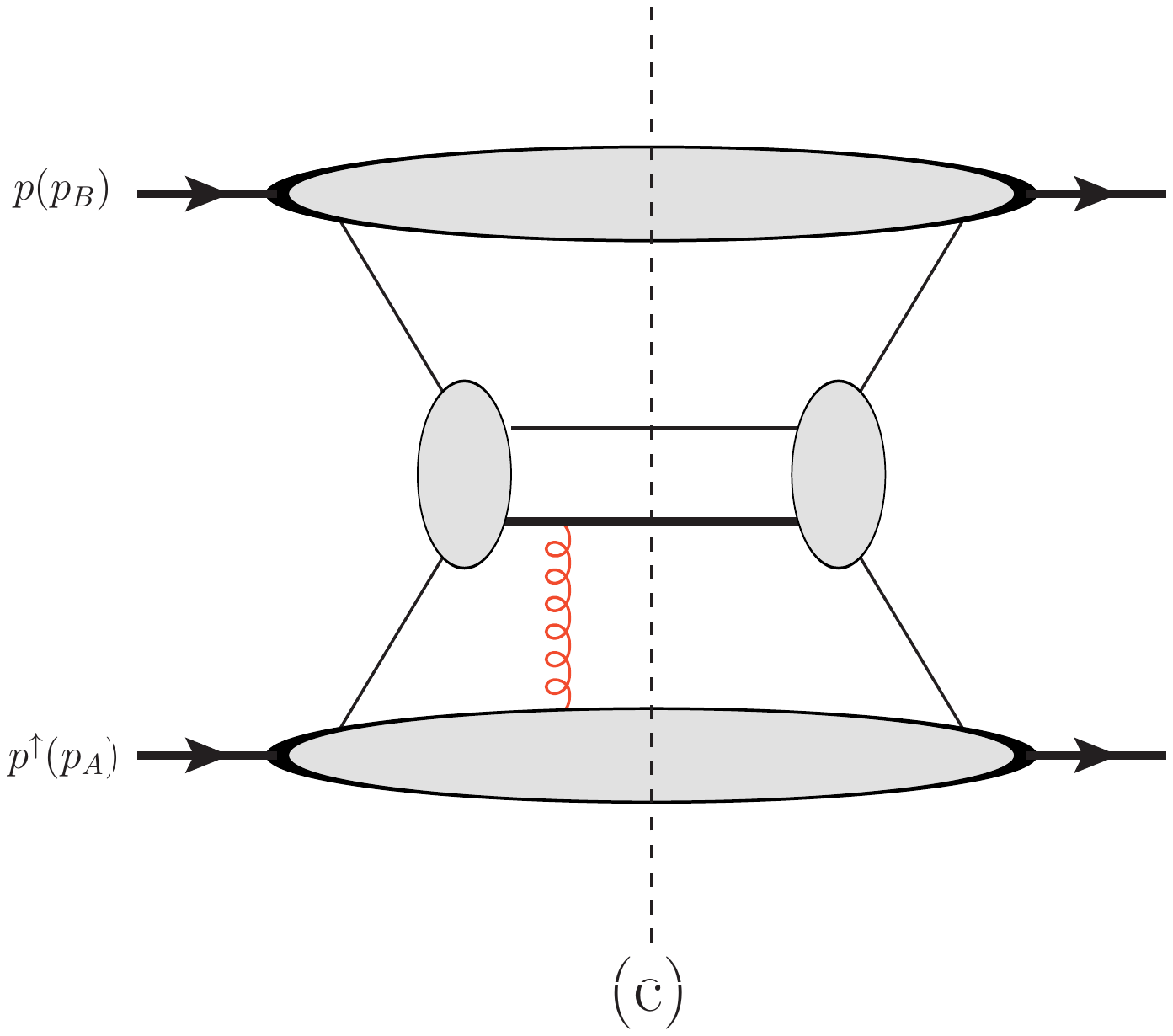}
\endminipage
 \advance\rightskip-1.5cm
 \vspace*{0.5cm}
	\caption{LO diagrams for $A^\uparrow+B\to C+X$ in (a) GPM, (b) CGI-GPM with initial-state interactions and (c) final-state interactions. The eikonal soft-gluon, shown in red, does not affect the kinematics of the process but only changes the colour flow.}
%   \end{adjustwidth}
	\label{fig:isfsdiagrams}
\end{figure}

In the Colour-Gauge Invariant GPM formalism, these ISIs and FSIs are treated at the one-gluon exchange level, i.e., by expanding the Wilson lines to leading order in the coupling constant, $g_s$. To illustrate this, in Fig.~\ref{fig:isfsdiagrams}, we show the diagrams for a process of the form $A^\uparrow +B\to C+X$ in (a) the standard GPM framework, and in the CGI-GPM framework with (b) the initial-state interactions and (c) the final-state interactions, both at one-gluon exchange level.  This approach was first proposed in Ref.~\cite{Gamberg:2010tj} where the effects of the ISI/FSIs were calculated for the various quark-initiated subprocesses (subprocesses with a quark from the polarised proton) involved in $p^\uparrow p\to\pi+X$. They used their formalism to reproduce the expected sign-flip between SIDIS and Drell-Yan Sivers asymmetries. 
 
In order to obtain the asymmetry in the CGI-GPM formalism, we need to take into account the effects of the ISIs and FSIs. For processes that probe the QSFs, this is done by making the following substitutions in Eqs.~3 and 9:
\be
 f^{\perp q}_{1T} H^U_{qb\to cd}\equiv f^{\perp q}_{1T}\sum_{i,j}\mathcal{A}^*_i\mathcal{A}_j\longrightarrow \sum_{i,j} \frac{C^{ij}_I+C^{ij}_{F_c}}{C^{ij}_U}f^{\perp q}_{1T} \mathcal{A}^*_i\mathcal{A}_j
\label{cgiqsf}
\ee
Note that we have used an alternative notation for the Sivers function, which is related to the one in Eqs.~\ref{directnum} and \ref{fragnum} by $f^{\perp q}_{1T}=-M_p~\Delta^N f_{i/p^\uparrow}/{2k_{\perp i}}$. The above expression has various terms which we will explain now: The $\mathcal{A}_i$ are the amplitudes for the different channels that contribute to the subprocess $qb\to cd$. Here, $q$ corresponds to the quark from the polarised proton. On the right hand side, $C_U^{ij}$ is the standard QCD colour factor for the product of amplitudes $\mathcal{A}^*_i\mathcal{A}_j$. $C^{ij}_I$ and $C^{ij}_{F_c}$ are colour factors for the diagrams with the initial state and final state interaction respectively.  Since the IS/FS interactions occur through eikonal soft-gluons, they do not affect the kinematics of the relevant diagrams and we can retain the same product of amplitudes $\mathcal{A}^*_i\mathcal{A}_j$, with the new modified colour factors, $(C^{ij}_I+C^{ij}_{F_c})/C^{ij}_U$ giving the appropriate colour flow.  Collecting the colour factors and the bilinear amplitudes in Eq.~\ref{cgiqsf}, we can 
define modified partonic hard parts,
\be
H^\text{mod}_{qb\to c d}\equiv \sum_{i,j} \frac{C^{ij}_I+C^{ij}_{F_c}}{C^{ij}_U} \mathcal{A}^*_i\mathcal{A}_j
\ee

While in Ref.~\cite{Gamberg:2010tj} only quark initiated subprocesses  were considered,  in Ref.~\cite{DAlesio:2017rzj} the CGI-GPM approach was extended to gluon initiated processes. They calculated the effects of the ISI/FSIs relevant for $p^\uparrow p\to J/\psi+X$ and $p^\uparrow p\to D+X$. For the case of gluons, the substitution required is,
\be
f^{\perp g}_{1T} H_U^{gb\to cd}\equiv f^{\perp g}_{1T}\sum_{i,j}\mathcal{A}^*_i\mathcal{A}_j\longrightarrow \sum_{i,j} \frac{C^{(f)ij}_I+C^{(f)ij}_{F_c}}{C^{ij}_U}f^{\perp g(f)}_{1T} \mathcal{A}^*_i\mathcal{A}_j +
 \frac{C^{(d)ij}_I+C^{(d)ij}_{F_c}}{C^{ij}_U}f^{\perp g(d)}_{1T} \mathcal{A}^*_i\mathcal{A}_j
\ee
Here things are different compared to the quark case since in the CGI-GPM framework, the process dependent gluon Sivers function can be written as a linear combination of two independent universal gluon distributions $f^{\perp g(f)}_{1T}$ and $f^{\perp g(d)}_{1T}$. These two gluon Sivers functions correspond to two possible ways of contracting the colour indices of the three gluon fields in the operator definition of the gluon Sivers function. The $f$-type denotes an completely antisymmetric contraction, $-if_{abc}$ and the $d$-type denotes a completely symmetric contraction, $d_{abc}$. We therefore define two modified hard parts, one associated with the $f$-type GSF and the other associated with the $d$-type GSF:
\be
H^{(f/d)}_{gb\to c d}\equiv \sum_{i,j} \frac{C^{(f/d)ij}_I+C^{(f/d)ij}_{F_c}}{C^{ij}_U} \mathcal{A}^*_i\mathcal{A}_j.
\ee

The two GSFs $f^{\perp g(f)}_{1T}$ and $f^{\perp g(d)}_{1T}$ have different properties. They have different behaviours under charge conjugation. $f^{\perp g(f)}_{1T}$ is $C$-even and  $f^{\perp g(d)}_{1T}$ is $C$-odd. Therefore only $f^{\perp g(f)}_{1T}$ is constrained by the Burkardt Sum Rule (BSR), which is defined in terms of $C$-even operators. We will be looking at bounds on the gluon contribution to $A_N$, from the BSR, both in the GPM and CGI-GPM frameworks. 
In this work, we calculate the effects of the initial and final state interactions for both direct photon production and photon production via fragmentation from quarks, following the techniques of  Ref.~\cite{DAlesio:2017rzj}.

\subsection{Modified hard-parts for direct photon production}
In direct photon production, there are no final state interactions since the partonic final state is a photon, which is colourless. The only partonic subprocesses which give access to the GSF at leading-order are $gq\to\gamma q$ and $gq\to \gamma \bar q$. {Here, the first parton to the left of the arrow in the subprocess label (in this case, the gluon) is the one coming from the polarised proton.} We have calculated the modified hard-parts for these subprocesses in the CGI-GPM framework.  While the two subprocesses have the same cross-section, they receive different modifications in the CGI-GPM framework due to their differing colour structures. The unpolarised and the modified hard-parts are given below:
\bea
H^U_{gq\to\gamma q}&=&-\frac{e_q^2}{N_c}\left[\frac{\hat u}{\hat s}+\frac{\hat s}{\hat u}\right]\\ \nonumber H^{(f)}_{gq\to\gamma q}&=& H^{(f)}_{g\bar q\to\gamma \bar q}=-\frac{1}{2}H^U_{gq\to\gamma q} \\ \nonumber H^{(d)}_{gq\to\gamma q}&=&-H^{(d)}_{g\bar q\to\gamma \bar q}=\frac{1}{2}H^U_{gq\to\gamma q} 
\eea
Let us note a few things: Firstly, the modified hard-parts are all half the magnitude of the unpolarised hard-part. Secondly the hard-part associated with the $f$-type GSF has a negative sign with respect to $H^U$ for processes with both quarks and antiquarks from the unpolarised proton, whereas the hard-part associated with the $d$-type GSF retains the same sign as $H^U$ for quarks and has a negative sign for antiquarks.

The quark Sivers functions also contribute to direct-photon production via the following subprocesses: $qg\to\gamma q$ and $q\bar q\to \gamma g$. The relevant hard-parts in this case are available in Ref.~\cite{Gamberg:2010tj} and we have reproduced them here for the sake of completeness. These are:
\bea
\label{qg}
H^U_{qg\to\gamma q}&=&-\frac{e_q^2}{N_c}\left(\frac{\hat t}{\hat s}+\frac{\hat s}{\hat t}\right) \\ \nonumber
H^\text{mod}_{qg\to \gamma q}=-H^\text{mod}_{\bar qg\to \gamma \bar q}&=&\frac{N_c}{N_c^2-1}~e_q^2\left(\frac{\hat t}{\hat s}+\frac{\hat s}{\hat t}\right) 
\eea

\bea
H^U_{q\bar q\to\gamma g}&=&\frac{N_c^2-1}{N_c^2}~e_q^2\left(\frac{\hat u}{\hat t}+\frac{\hat t}{\hat u}\right) \\ \nonumber
H^\text{mod}_{q\bar q\to\gamma g}=-H^\text{mod}_{\bar q q\to\gamma g}&=&\frac{e_q^2}{N_c^2}\left(\frac{\hat u}{\hat t}+\frac{\hat t}{\hat u}\right) 
\eea
Note that for the process $qg\to\gamma q$ the modified hard-part is roughly similar in size and has a negative sign for quarks and positive sign for antiquarks (relative to $H^U$). For the process $q\bar q\to\gamma g$ the modified hard-part is {\it much} smaller (by a factor of 8) and has a positive sign for quarks and a negative sign for antiquarks (relative to $H^U$). As we will see, this has significant consequences for the SSAs. %Since the overall $q\bar q\to \gamma g$ cross-section is small in the kinematic region we are interested in it, any asymmetry due to this process will be highly suppressed in the CGI-GPM framework and the only relevant processes are $q g\to\gamma g$ and $\bar q g\to \gamma g$, which have hard-parts with opposite signs (Eq.~\ref{qg}).
\\[6ex]
\subsection{Modified hard parts for photon production from fragmentation}
For photon production via fragmentation at leading order, there are seven processes that give access to the GSF: $gq\to qg$, $g\bar q\to \bar q g$, $gq\to gq$, $g\bar q\to  g \bar q$, $gg\to gg$ , $gg\to q\bar q$, and $g g\to \bar q q$. The modified hard-parts for the first five of these processes are not available in literature and we have calculated them. We first give the hard-parts for $gq\to qg$ and $g\bar q\to \bar q g$:
\\
\bea
H^U_{gq\to qg}&=&- \frac{(\shat^2+\that^2)}{2~\shat~\that~\uhat^2}\left[\shat^2+\that^2-\frac{\uhat^2}{N_c^2}\right] \\[3ex]
\nonumber
H^{(f)}_{gq\to qg}=H^{(f)}_{g\bar q\to \bar q g}&=&- \frac{(\shat^2+\that^2)}{4~\shat~\that~\uhat^2}\left[2~\that~\uhat+\uhat^2\right]    \\[1ex]
\nonumber
H^{(d)}_{gq\to qg}=-H^{(d)}_{g\bar q\to \bar q g}&=& - \frac{(\shat^2+\that^2)}{4~\shat~\that~\uhat^2}\left[\shat^2+\that^2-2~\frac{\uhat^2}{N_c^2}\right]
\eea
\\
The hard-parts for $gq\to gq$ and $g\bar q\to  g \bar q$ are as follows:
\\
\bea
H^U_{gq\to gq}&=&- \frac{(\shat^2+\uhat^2)}{2~\shat~\that^2~\uhat}\left[\shat^2+\uhat^2-\frac{\that^2}{N_c^2}\right] \\[3ex]
\nonumber
H^{(f)}_{gq\to gq}=H^{(f)}_{g\bar q\to  g\bar q}&=&- \frac{(\shat^2+\uhat^2)}{4~\shat~\that^2~\uhat}\left[\shat^2+\frac{\that^2}{N_c^2}\right] \\[1ex]
\nonumber
H^{(d)}_{gq\to gq}=-H^{(d)}_{g\bar q\to  g\bar q}&=&+ \frac{(\shat^2+\uhat^2)}{4~\shat~\that^2~\uhat}\left[\shat^2-2~\uhat^2+\frac{\that^2}{N_c^2}\right]
\eea
\\
The hard-parts for $gg\to gg$ are:
\\
\bea
H^U_{gg\to gg}&=& \frac{4N_c^2}{N_c^2-1}\frac{(\that^2+\that~\uhat+\uhat^2)^3}{\shat^2~\that^2~\uhat^2} \\[3ex]
\nonumber
H^{(f)}_{gg\to gg}&=&\frac{N_c^2}{N_c^2-1}\frac{(\that^2+\that~\uhat+\uhat^2)^2}{\shat^2~\that^2~\uhat^2}~(2~\that~\uhat+\uhat^2) \\[1ex]
\nonumber
H^{(d)}_{gg\to gg}&=&0
\eea

Finally, the hard-parts for $gg\to q\bar q$ and $gg\to \bar q q$ were calculated in Ref.~\cite{DAlesio:2017rzj} and are reproduced below for sake of completeness:
\\
\bea
H^U_{gg\to q\bar q}&=&\frac{N_c}{N_c^2-1}~\frac{1}{\that~\uhat}\left(\frac{N_c^2-1}{2N_c^2}-\frac{\that~\uhat}{\shat^2}\right)(\that^2+\uhat^2) \\[1ex]
\nonumber
H^{(f)}_{gg\to q\bar q}=H^{(f)}_{gg\to \bar q q}&=&-\frac{N_c}{4(N_c^2-1)}~\frac{1}{\that~\uhat}~\left(\frac{\that^2}{\shat^2}+\frac{1}{N_c^2}\right)(\that^2+\uhat^2) \\[1ex]
\nonumber
H^{(d)}_{gg\to q\bar q}=-H^{(d)}_{gg\to \bar q q}&=&-\frac{N_c}{4(N_c^2-1)}~\frac{1}{\that~\uhat}~\left(\frac{\that^2-2~\uhat^2}{\shat^2}+\frac{1}{N_c^2}\right)(\that^2+\uhat^2)
\eea

The relevant hard-parts for quark initiated subprocesses that give access to the QSFs can be found in Ref.~\cite{Gamberg:2010tj}. We do not present them here.

\section{Parametrisation of the TMDs}
In this section we give the details of the functional forms and parameters that we use for the TMDs. For the unpolarised TMDs we adopt the commonly used form with the collinear PDF multiplied by a Gaussian transverse momentum dependence,
\be
f_{i/p}(x,k_\perp;Q)=f_{i/p}(x,Q)\frac{1}{\pi\langle k_\perp^2\rangle}e^{-k_\perp^2/\langle k_\perp^2\rangle}
\ee
with $\langle k_\perp^2\rangle=0.25\text{ GeV}^2$. 
As with the unpolarised densities, we use a similar factorised Gaussian form for the photon fragmentation function,
\be
D_{\gamma/c}(z,\bfk_\gamma)=D_{\gamma/c}(z)\frac{1}{\pi\langle k^2_{\perp \gamma}\rangle}e^{-k_\gamma^2/\langle k^2_{\perp \gamma}\rangle}
\ee
with $\langle k^2_{\perp \gamma}\rangle=0.25$ GeV$^2$.

Since we give predictions using the GSF fits of Ref.~\cite{DAlesio:2015fwo}, we adopt the functional form of the Sivers functions used therein:
\be
\Delta ^N f_{i/\pup}(x,k_{\perp};Q)=2\mathcal{N}_i(x)f_{i/p}(x,Q)\frac{\sqrt{2e}}{\pi}\sqrt{\frac{1-\rho}{\rho}}k_\perp \frac{e^{-k^2_\perp/\rho\langle k^2_\perp\rangle}}{\langle k^2_\perp\rangle^{3/2}}
\ee
with $0<\rho<1$. Here $\mathcal{N}_i(x)$ parametrises the $x$-dependence of the Sivers function and is generally written as
\be
\mathcal{N}_i(x)=N_gx^{\alpha_i}(1-x)^{\beta_i}\frac{(\alpha_i+\beta_i)^{\alpha_i+\beta_i}}{\alpha_i^{\alpha_i}\beta_i^{\beta_i}}
\ee
For the  Sivers function to satisfy the positivity bound,
\be
\frac{|\Delta^Nf_{i/p^\uparrow}(x,\mathbf{k}_\perp)|}{2f_{i/p}(x,\mathbf{k}_\perp)}\leq 1\>\forall \>x, \mathbf{k}_\perp,
\label{positivity}
\ee
it is necessary to have  $|\mathcal{N}_i(x)|<1$.

In order to study the efficacy of the probe, we explore the following choices for the Sivers functions:
\begin{enumerate}
\item Quark and gluon Sivers functions with the positivity bound saturated, viz. $\mathcal{N}_i(x)=1$ and $\rho=2/3$.
\item The SIDIS1~\cite{Anselmino:2005ea} and SIDIS2~\cite{Anselmino:2008sga} fits of the QSFs, along with the associated GSF fits from Ref.~\cite{DAlesio:2015fwo}.
\end{enumerate}

The first choice, which we will refer to as `saturated' Sivers function is the maximal Sivers function allowed by the positivity bound for a fixed width $\langle k_\perp^2 \rangle$ and $\rho$, with a particular choice of unpolarised collinear gluon density. The parameter $\rho$ is set to 2/3 in order to maximise the first $k_\perp$-moment of the Sivers function, following Ref.~\cite{DAlesio:2010sag}. Using the saturated Sivers functions for quarks and gluons allows us to study the general kinematic dependencies of the asymmetry and the relative importances of the quark and gluon contributions to the asymmetry. It also lets us study how uncertainties in the knowledge of the collinear, unpolarised gluon and sea quark densities might impact the probe.

SIDIS1~\cite{Anselmino:2005ea} and SIDIS2~\cite{Anselmino:2008sga} are two different sets of QSFs both fitted to data on single-spin asymmetry in SIDIS. The SIDIS1 set was fit to data on pion production at HERMES and flavour unsegregated data on positive hadron production at COMPASS. They used quark fragmentation functions by Kretzer~\cite{Kretzer:2000yf}. The data, being flavour unsegregated, were not sensitive to sea quark contributions. Hence this set contains parametrisations for only $u$ and $d$ quark Sivers functions. The SIDIS2 set was fit to flavour-segregated data on hadron production at HERMES and COMPASS. Since strange meson production receives contribution from sea quarks, this fit includes sea quark Sivers functions as well. Further they used fragmentation functions by de Florian, Sassot and Stratmann (DSS)~\cite{deFlorian:2007aj} in this second fit. 

\begin{table}[t]
\begin{center}
\begin{tabular}{l l l l }
\hline
\hline
\noalign{\vspace{1pt}}
%\multicolumn{3}{|c|}{~~~}\\
\multicolumn{3}{ c }{~~~~~~~~~~~~~~~~~~SIDIS2 QSFs from Ref.~\cite{Anselmino:2008sga}~~~~~~~~~~~~~~~~~~~~}\\
%\multicolumn{3}{|c|}{~~~}\\
\noalign{\vspace{1pt}}
\hline
%~&~&~\\
~~~$N_{u} = 0.35 $ \hspace*{1cm} &
~~~~~$N_{d} = -0.90 $ &
~~~$N_{s} = -0.24 $~~ \\
~~~$N_{\bar u} =  0.04 $&
~~~~~$N_{\bar d} =  -0.40 $&
~~~$N_{\bar s} =  1 $ \\
~~~$\alpha _u = 0.73 $ &
~~~~~$\alpha_d = 1.08 $  &
~~~$\alpha_{sea} = 0.79 $ \\
~~~$\beta \;\;= 3.46$ &
~~~~~$M_1^2 = 0.34 $ GeV$^2$~  &
~~~~~ \\
%~&~&~\\
%
\hline
\hline
\noalign{\vspace{1pt}}
%\multicolumn{3}{|c|}{~~~}\\
\multicolumn{3}{ c }{~~~~~~~~~~~~~~~~~~SIDIS1 QSFs from Ref.~\cite{Anselmino:2005ea}~~~~~~~~~~~~~~~~~~~~}\\
%\multicolumn{3}{|c|}{~~~}\\
\noalign{\vspace{1pt}}
\hline
%~&~&~\\
~~~$N_{u} = 0.32  $ \hspace*{1cm} &
~~~~~$N_{d} = -1.00 $ &
~~~~~ \\
~~~$\alpha _u = 0.73 $ &
~~~~~$\alpha_d = 1.08 $  &
$M_0^2 = 0.32 $ GeV$^2$ \\
~~~$\beta_u=0.53$ &
~~~~~$\beta_d=3.77$~  &
~~~~~ \\
%~&~&~\\
%
\hline
\hline
\noalign{\vspace{1pt}}
\multicolumn{4}{ c }{~~~~~~~~~~~~~~~~~~SIDIS2 GSF from Ref.~\cite{DAlesio:2015fwo}~~~~~~~~~~~~~~~~~~~~}\\
\noalign{\vspace{1pt}}
\hline
%~&~&~\\
$N_{g} = 0.05$ &
$\alpha_g=0.8 $ &
$\beta_g=1.4$ &$\rho=0.576$\\
\hline
\hline
\noalign{\vspace{1pt}}
\multicolumn{4}{ c }{~~~~~~~~~~~~~~~~~~SIDIS1 GSF from Ref.~\cite{DAlesio:2015fwo}~~~~~~~~~~~~~~~~~~~~}\\
\noalign{\vspace{1pt}}
\hline
%~&~&~\\
$N_{g} = 0.65$ &
$\alpha_g=2.8 $ &
$\beta_g=2.8$ &$\rho=0.687$\\
\hline
\hline
\end{tabular}
\end{center}
\caption{
Parameters for the various Sivers function fits used.}

\end{table}

Associated with these two QSF sets are the two fits of the GSF from Ref.~\cite{DAlesio:2015fwo}, which we will refer to as SIDIS1 and SIDIS2 as well. These were both obtained by constraining the gluon contribution to data on $A_N$ in midrapidity pion production at RHIC~\cite{Adare:2013ekj}, with the aforementioned QSFs being used to account for the quark contribution. While both the GSFs --- along with their associated QSF sets --- give good fits to the data on midrapidity pion production, they have very different $x$-dependencies. SIDIS1 is larger in the moderate-$x$ region and SIDIS2 is larger in the low-$x$ region. The values of the parameters of the two QSF sets as well as the associated GSF fits are given in Table I.

\section{Numerical Estimates for Direct Photon Production}

In this work we wish to see if prompt-photon production at RHIC can be used to obtain any information on the Gluon Sivers function and whether it has any discriminating power between  the available GSF parametrisations. To this end we need to consider various things such as the kinematics, contribution of QSFs to the asymmetry, uncertainties in the knowledge of the collinear, unpolarised PDFs, existing bound on the GSFs etc. 
\begin{figure}[t]
\vspace*{-1cm}
\begin{center}
\includegraphics[width=\linewidth]{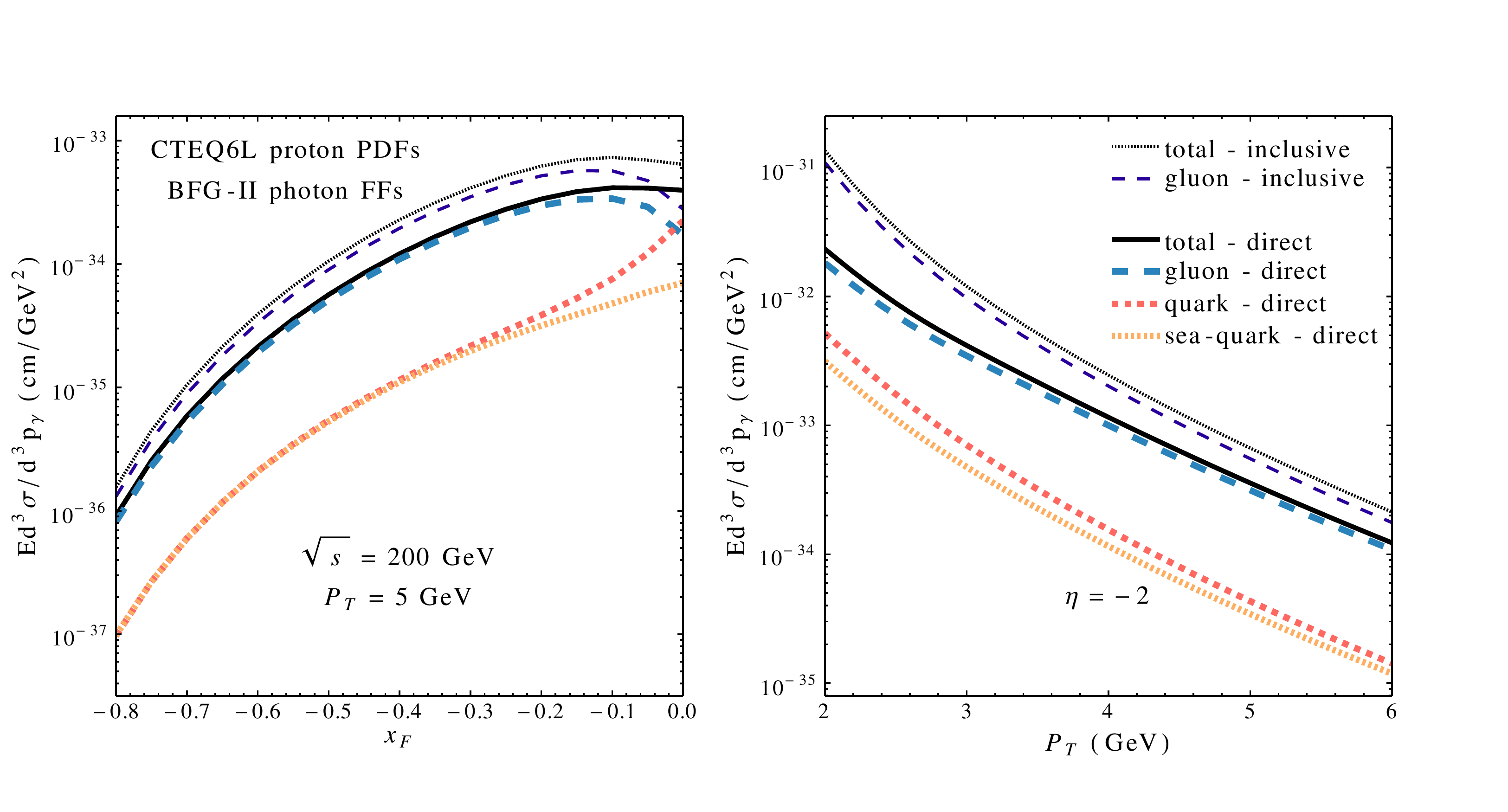}
\vspace*{-1.5cm}
\caption{Unpolarised Lorentz-invariant cross-section for prompt photon production at RHIC ($\sqrt{s}=200$ GeV) as a function of $x_F$ (at $P_T=5$ GeV, left panel) and $P_T$ (at rapidity $\eta=-2$, right panel). The thick lines indicate the direct photon contributions and the thin lines indicate the inclusive (direct and fragmentation) prompt photon contributions. The renormalisation and factorisation scales were chosen to be $Q=P_T$.}
\label{logcs}
\end{center}
\end{figure}

Direct-photon production at $x_F<0$ was analysed in Ref.~\cite{Schmidt:2005gv}. The authors noted that this region is dominated by the Compton scattering process $gq\to\gamma q$ with the  gluons coming from the transversely polarised proton. Hence $x_F<0$ had been identified as the region appropriate for probing the GSF. They  had then suggested in general, the large-$P_T$ region in the backward hemisphere as the region to be used.  In Fig.~\ref{logcs}, we show the unpolarised Lorentz-invariant cross-section for the production of prompt photons at RHIC energy, $\sqrt{s}=200$ GeV, as a function of $x_F$, at fixed $P_T=5$ GeV (left panel) and as a function of $P_T$ at a fixed pseudorapidty $\eta=-2$. By prompt photons, we mean both direct and fragmentation photons. The plot also shows the direct photon contribution separately. The inclusive components are given as thin lines and the direct components are given as thick lines.  For both the inclusive and direct components, the contribution from subprocesses with the gluon in the transversely polarised proton is indicated separately. Further for the direct component, the contributions from subprocesses with quarks, as well as sea quarks in the transversely polarised proton are indicated separately. In obtaining these numbers, we used CTEQ6L~\cite{Pumplin:2002vw} PDFs  for the collinear part of the proton densities and the BFGII~\cite{Bourhis:1997yu} parton-to-photon FFs for the collinear part of the fragmentation functions. A Gaussian width $\langle k_\perp^2\rangle=0.25\text{ GeV}^2$ was used for both gluon and quark TMD-PDFs as well as for  the TMD photon FFs. The renormalisation and factorisation scales were chosen to be $Q=P_T$.

As we can see from Fig.~\ref{logcs}, the total prompt photon cross-section inclusive of direct and fragmentation photons is of the same order of magnitude as the direct photon cross-section alone. Overall, gluons dominate the production process in the kinematic regions we consider. Of the direct photon component, for $P_T\gtrsim3$ GeV and $x_F\lesssim-0.1$, more than 75\% of the cross-section is from the Compton process $gq\to\gamma q$ with the initial state gluon from the forward-going proton. Among  the contributions to direct photons coming from quark-initiated processes (processes with quarks in the transversely polarised proton), the sea-quarks  give  the dominant component, being around $10$-$15\%$ of the total cross-section. In the low $P_T$ ($\lesssim3$ GeV) and $x_F>-0.1$, regions where the cross-section is highest, the contributions from gluons, valence quarks and sea quarks are similar in magnitude, hence precise knowledge of the collinear densities will be required for any analysis of the SSA in this kinematic region. The qualitative details of the above observations remain unchanged even when we use other available fits of the collinear densities.

\begin{figure}[t]
\vspace*{-1cm}
\begin{center}
\includegraphics[width=\linewidth]{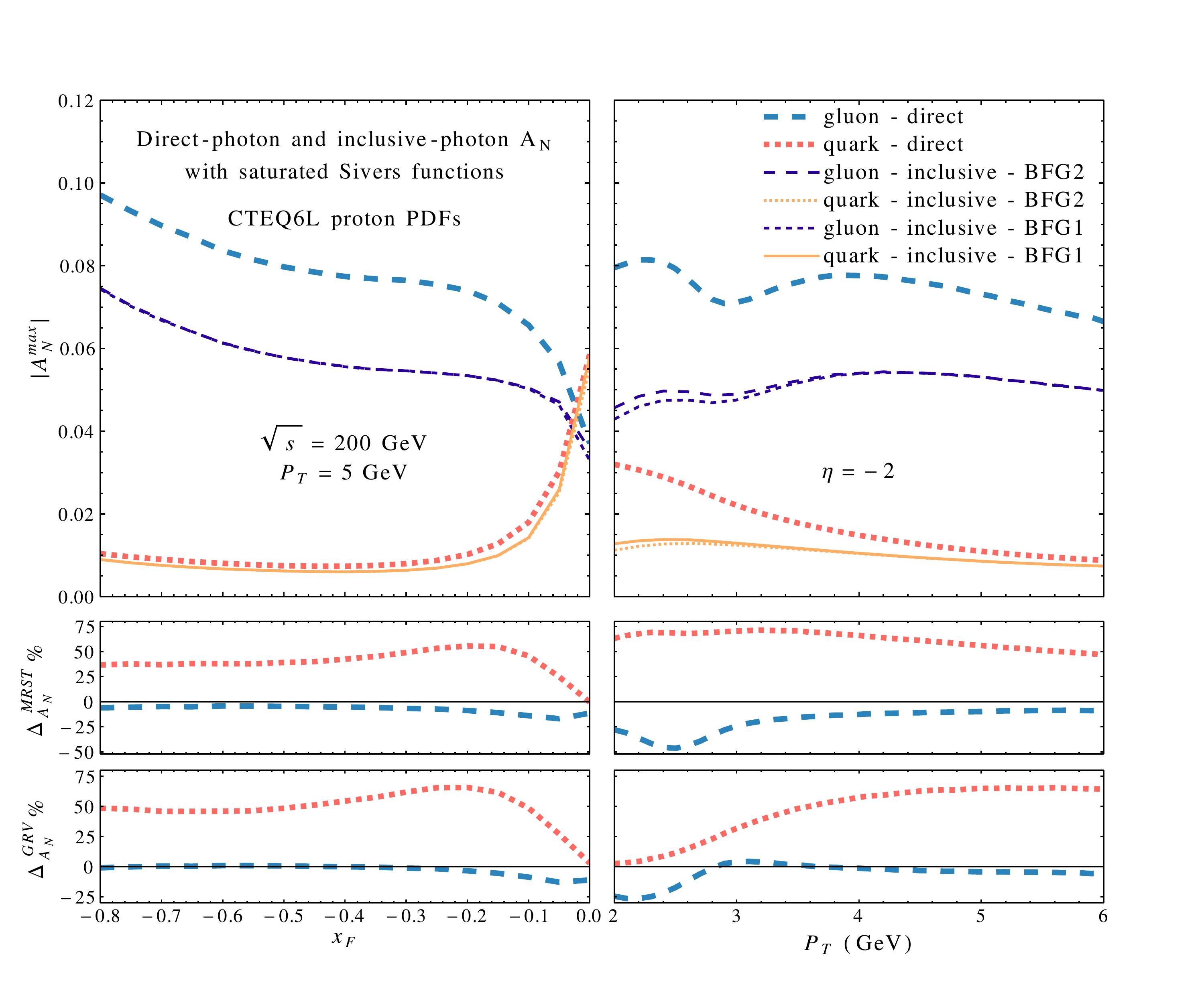}
\vspace*{-1.5cm}
\caption{SSA in prompt photon production using saturated quark and gluon Sivers functions. Results shown as a function of $x_F$ (at $P_T=5$ GeV, left panel) and $P_T$ (at rapidity $\eta=-2$, right panel). Thick lines indicate asymmetry in direct photons and thin lines indicate asymmetry in all prompt photons inclusive of fragmentation photons. The top pair of panels show the results obtained using CTEQ6L~\cite{Pumplin:2002vw} PDFs. The pair of panels below them indicate the percentage change in the direct photon results when MRST2001LO~\cite{Martin:2002dr} PDFs are used, i.e., $100\times (A_N^\text{CTEQ}-A_N^\text{MRST})/A_N^\text{MRST}$. The pair of panels further below show the percentage change in direct photon results when GRV98LO~\cite{Gluck:1998xa} PDFs are used.}
\label{gpmsatan}
\end{center}
\end{figure}

\subsection{GPM}

\subsubsection{Asymmetry estimates using saturated Sivers functions}

We now consider results for the asymmetry obtained using saturated Sivers functions for gluons and quarks. As mentioned in Sec.~IV, the saturated Sivers function can be taken to be an upper bound on the Sivers function as allowed by the positivity bound, Eq.~\ref{positivity}. Considering the asymmetries obtained using the saturated Sivers functions is useful for two reasons: First it allows us to see the maximum possible sizes of the effect and  consider the relative importances of the valence quarks, sea quarks and gluons in the absence of any inputs from fits of the QSFs and the GSF. Second, by considering different choices of collinear densities we can get an idea of how uncertainties in the knowledge of the collinear PDFs can affect the analysis. We will consider both now.

In Fig.~\ref{gpmsatan}, we show the asymmetries obtained in the GPM framework using saturated Sivers functions for the gluon and all flavours of quarks. A positive sign is used for QSFs of all flavours. We show the asymmetry in the direct photons alone, as well as the asymmetry inclusive of direct and fragmentation photons. The inclusive photon asymmetries are given as thin lines and the direct photon asymmetries are given as thick lines. The gluon and quark contributions for both are shown separately. In the bottom panels, we show the percentage change in direct photon results when using different two different choices of LO collinear densities, MRST2001LO~\cite{Martin:2002dr} and GRV98LO~\cite{Gluck:1998xa}.

We first discuss the direct photon results. When using CTEQ6L PDFs, the saturated GSF (with QSFs set to zero) gives an asymmetry of upto almost 10\% at $x_F=-0.8$ and $8\%$ at $P_T=2$ GeV. The asymmetry from the saturated QSFs is largest at low values of $|x_F|$ and $P_T$ being 6\% at $x_F=0$ and $3\%$ at $P_T=2$ GeV. It can be seen from the plot of the cross-section in Fig.~\ref{logcs} that the contribution of sea quarks is significant in the entire kinematics range and is in fact dominat for $x_F<-0.2$ at $P_T=5$ GeV.

From the bottom two sets of panels of Fig.~\ref{gpmsatan}, we can see how the direct photon results vary when using the MRST2001LO and GRV98LO PDF sets for the collinear parts of the TMDs. This gives us an idea of how uncertainties in our knowledge of the collinear densities can impact the predictions of SSA in the considered kinematic region. Both MRST and GRV give quark contributions that are larger by $50$-$70\%$, whereas gluon contributions are mostly similar throughout the kinematic range except at low-$P_T$ values at $\eta=2$ where MRST gives a gluon contribution that is smaller by upto 50\% and GRV give a gluon contribution that is smaller by upto 25\%.

%\sout{Finally we look at the impact that fragmentation contributions can have on the asymmetry. We have plotted the GSF-only and QSF-only contributions of inclusive asymmetry for two different choices of parton-to-photon FFs, BFGI and BFGII. \textcolor{red}{\sout{As mentioned in Sec.~II, the contribution from the fragmentation part can be reduced by using a photon isolation cut but cannot be completely eliminated.}} \textcolor{red}{\sout{Therefore it}} \textcolor{blue}{Since it} is important to get an idea of the maximum impact that fragmentation photons can have on the asymmetry\textcolor{red}{\sout{. To this end}}, we \textcolor{red}{\sout{need to}} look at the asymmtery inclusive of both direct and fragmentation photons. As can be seen from the thin violet curves in Fig.~\ref{gpmsatan}, the GSF contribution of the inclusive asymmetry is diluted by anywhere between 10-50\% of the direct-photon value. The QSF contribution to the inclusive asymmetry also has the same fate. Hence, it is important to remove the fragmentation contribution as much as possible. Overall the results for the inclusive asymmetry do not depend much on the choice of the FF set used. Results  obtained for the inclusive asymmetry for different choices of collinear PDFs also  show similar trend.}

Finally we consider the fragmentation contributions. Since it is important to get an idea of the maximum impact that fragmentation photons can have on the asymmetry, we look at the asymmtery inclusive of both direct and fragmentation photons.  We have plotted the GSF-only and QSF-only contributions of inclusive asymmetry for two different choices of parton-to-photon FFs, BFGI and BFGII. As can be seen from the thin violet curves in Fig.~\ref{gpmsatan}, the GSF contribution of the inclusive asymmetry is diluted by anywhere between 10-50\% of the direct-photon value. The QSF contribution to the inclusive asymmetry also has the same fate. Hence, it is important to remove the fragmentation contribution as much as possible. Overall the results for the inclusive asymmetry do not depend much on the choice of the FF set used. Results  obtained for the inclusive asymmetry for different choices of collinear PDFs also  show a similar trend.

\subsubsection{Asymmetry estimates using existing fits as well as constraints from the Burkardt Sum Rule}

We now consider existing information on the gluon and quark Sivers functions that have been obtained from fits to data. As mentioned in Sec.~IV, we consider two different sets of the QSFs, labelled SIDIS2 and SIDIS1, along with their associated GSF fits from Ref.~\cite{DAlesio:2015fwo}. Apart from the two available fits of the GSF, a general indirect bound on the GSF can be obtained based on the Burkardt Sum Rule~\cite{Burkardt:2004ur}. This sum rule for the Sivers functions is essentially the requirement that the net transverse momentum of all the  partons in a transversely polarised proton must vanish,
\be
\langle \bfk_\perp\rangle=\sum_{i=q,\bar q,g}\langle\bfk_{\perp i}\rangle=\sum_{i=q,\bar q,g}\int  dx f^{\perp(1)i}_{1T}(x)=0
\ee
where $f^{\perp(1)i}_{1T}$ is the first transverse moment of the Sivers function,
\be
f^{\perp(1)i}_{1T}\equiv-\int d^2\bfk_\perp\frac{k_\perp}{4M_p}\Delta^Nf_{g/p^\uparrow}(x,k_\perp).
\ee

The QSF fits of the SIDIS2 set, taken with their associated errors, allow the gluon contribution to be in the following range:
\be
-10\leq\langle k_{\perp g}\rangle\leq48\text{ MeV}
\label{kTbound}
\ee
This is to be compared with the following values obtained for the quarks in the SIDIS2 fit:
\be
\langle k_{\perp u}\rangle=98^{+60}_{-28}\text{ MeV, }\hspace*{0.5cm}\langle k_{\perp d}\rangle=-113^{+45}_{-51}\text{ MeV}
\ee
Since this is an indirect constraint based on a quantity integrated over the parton light-cone momentum fraction $x$, this tells us nothing about the $x$-dependence --- or for that matter the $k_\perp$-dependence --- of the GSF. However, we can still get an idea of the possible sizes of the gluon contibution to the asymmetry in direct photon production by considering various possible GSF parameter sets, $N_g$, $\alpha_g$, $\beta_g$ and $\rho$, that result in Sivers function obeying Eq.~\ref{kTbound}. We obtained these sets by performing a scan over the parameter space, varying the parameters in the following ranges: $N_g$ in the range -1 to 1 in steps of 0.1, $\alpha_g$ and $\beta_g$ in the range 0 to 4 in steps of 0.2, and $\rho$ in the range 0.1 to 0.9 in steps of 0.05. This exercise is justified given the very different $x$-dependencies of the SIDIS1 and SIDIS2 fits of the GSF, both of which were obtained using QSF sets that describe SIDIS data equally  well.

\begin{figure}[t]
\vspace*{-1cm}
\begin{center}
\includegraphics[width=\linewidth]{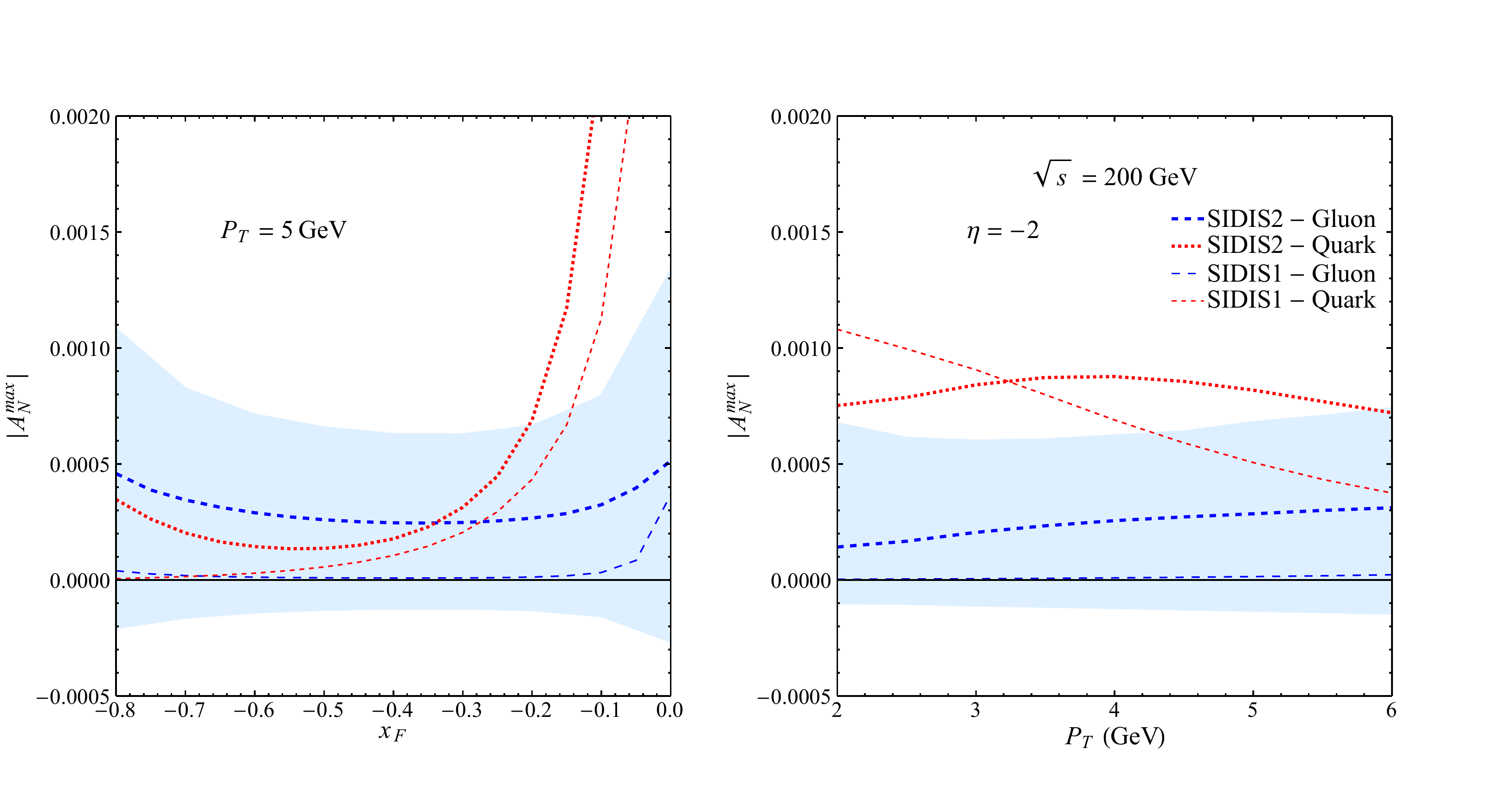}
\vspace*{-1.5cm}
\caption{Burkardt Sum Rule based constraints on $A_N$ in the GPM framework and predictions of asymmetry using the SIDIS2~\cite{Anselmino:2008sga} and SIDIS1~\cite{Anselmino:2005ea} QSFs and associated GSFs~\cite{DAlesio:2015fwo}. The light blue band shows the envelope of asymmetries obtained from GSFs that obey the BSR based constaint, Eq.~23. }
\label{GPMANband}
\end{center}
\end{figure}

In Fig.~\ref{GPMANband} we plot the band of direct photon $A_N$ values obtained with the GSF parameter sets results from the scan. This is shown by the light-blue shaded region. Along with it, we also plot the quark and gluon contribution to the direct photon asymmetry as given by the SIDIS2 and SIDIS1 sets. For the collinear parts of the densities, we used GRV98LO PDFs as they were used in the extraction of the SIDIS2 fits and the BSR bound. Both the GSF fits give asymmetries that lie well within the asymmetry band given by the constraint in Eq.~\ref{kTbound}. The indirect bound allows a gluon contribution to the asymmetry of upto 0.07-0.1\%. Overall we find that both the SIDIS1 and SIDIS2 GSF fits, as well as the Burkardt Sum Rule constraints predict negligible values for the gluon contribution to the asymmetry.

\subsection{CGI-GPM}
\subsubsection{Asymmetry estimates using saturated Sivers functions}
We now consider the probe in the context of the Colour-Gauge Invariant generalised parton model.

\begin{figure}[t]
\vspace*{-1cm}
\begin{center}
\includegraphics[width=\linewidth]{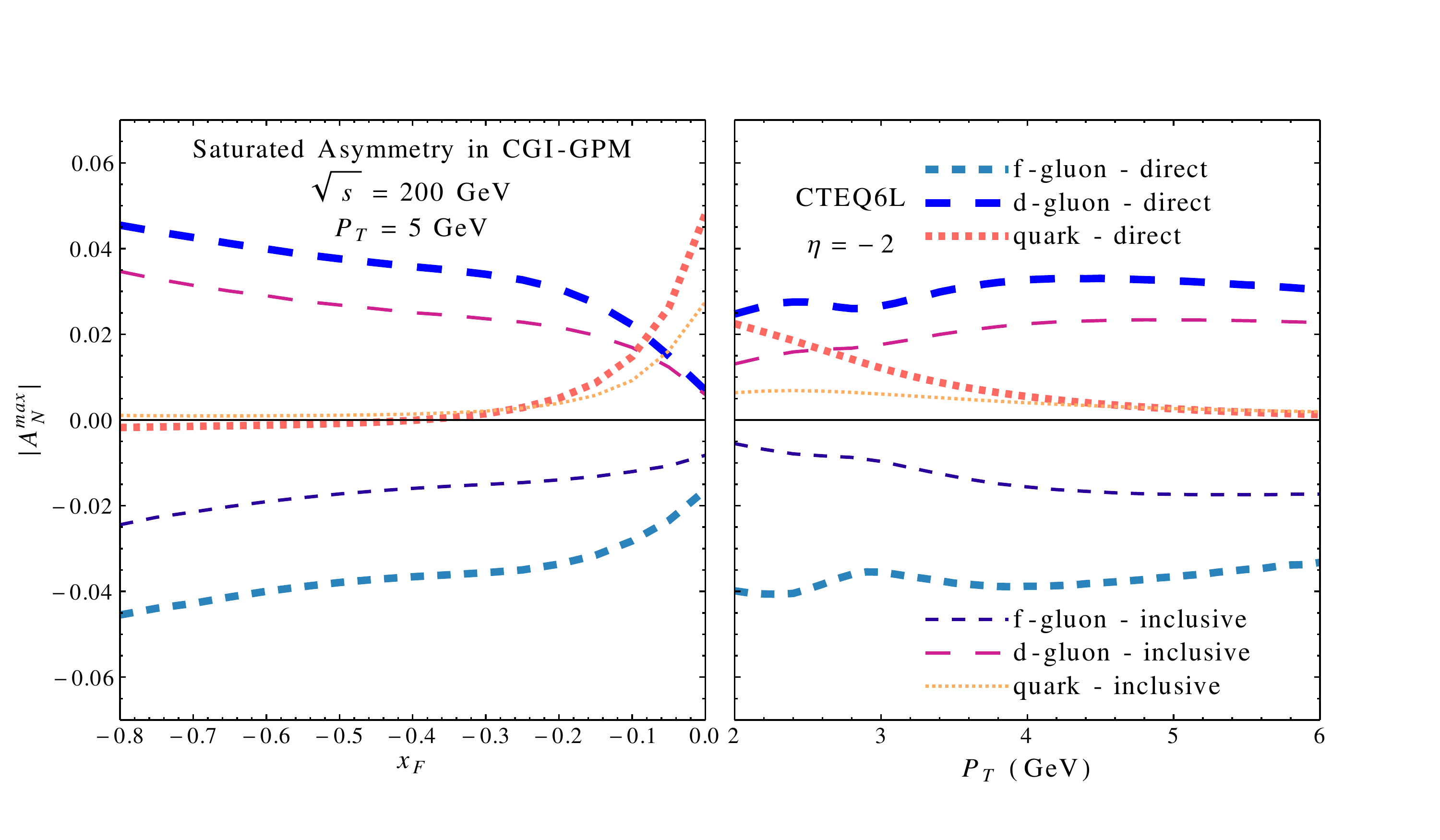}
\vspace*{-1.5cm}
\caption{SSA in direct-photon production using the CGI-GPM framework with saturated quark and gluon Sivers functions. Results shown as a function of $x_F$ (at $P_T=5$ GeV, left panel) and $P_T$ (at rapidity $\eta=-2$, right panel). Note that direct photon asymmetries from $f$ and $d$ type GSFs are shown with opposite signs. This is because their modified hard-parts associated with them have opposite signs. Depending on the relative signs of the two GSFs and also their relative magnitudes, the contributions from the two GSFs may add up or cancel out. Results obtained using CTEQ6L PDFs.}
\label{cgigpmsatan}
\end{center}
\end{figure}

In Fig.~\ref{cgigpmsatan}, we show the asymmetries obtained using saturated Sivers functions, in the CGI-GPM framework. We show the asymmetry in the direct photons alone, as well as the asymmetry inclusive of direct and fragmentation photons. 

We first discuss the direct photon results. We can see that $f$-type contribution has a negative sign. This is beacuse the hard-parts $H^{(f)}_{gq(\bar q)\to\gamma q(\bar q)}$ have an opposite sign with respect to $H^U_{gq\to\gamma q}$, whereas the hard-parts $H^{(d)}_{gq(\bar q)\to\gamma q(\bar q)}$ have the same sign as $H^U_{gq\to\gamma q}$. Since the modified hard-parts associated with the two GSFs are half the magnitude of the unpolarised hard-part, the asymmetry estimates will be halved in magnitude as compared to the GPM result. Further the $d$-type hard-parts have opposite signs for quarks and antiquarks, i.e., $H^{(d)}_{gq\to\gamma q}=-H^{(d)}_{g\bar q\to\gamma \bar q}$, so in regions where the antiquark content of the unpolarised proton is significant, there is a further suppression for the $d$-type contribution. This can be clearly seem from the panel for $\eta=-2$. Overall,  depending on the relative signs of the two GSFs and also their relative magnitudes, the contributions from the two of them may add up or cancel out.

Unlike the gluon contribution which decreases by a factor of two with respect to the GPM, the changes in the quark contribution are not so straightforward. We should first note that in obtaining the above plot, the signs of the saturated $u$, $d$ and $s$ quark Sivers functions were chosen to be negative, i.e., $\mathcal{N}_u(x)=\mathcal{N}_d(x)=\mathcal{N}_s(x)=-1$. The modified hard-part for the process $qg\to\gamma q$ has a negative sign with respect to the unpolarised hard-part, therefore the negative sign ensures that the resulting quark contribution to the asymmetry is positive in regions where $qg\to\gamma q$ dominates over $q\bar q\to\gamma g$. This is the case in the neighbourhood of midrapidity, $x_F>-0.2$ . For further backward regions, the process $q\bar q\to\gamma g$ is dominant and the quark asymmetries are highly suppressed since the modified hard-part for this process is suppressed by a factor of eight with respect to the unpolarised hard-part (c.f.~Eq.~15). 

Overall for the case of direct photons, the saturated Sivers function based analysis in the CGI-GPM framework leads us to the following conclusions on how things are different with respect to the GPM: Both the $f$ and $d$ type GSF contributions to the asymmetry are about half the magnitude of the GPM result. If the two GSFs are similar in magnitude, their overall contributions may add up or cancel each other out depending on their signs, which are unknown.  The changes in the quark contributions depend on the various signs involved. In highly backward regions, the quark contribution to the asymmetry is highly suppressed with respect to the gluon contribution. This is because a) the quark contribution happens through the $q\bar q\to \gamma g$ subprocess which has a much smaller cross-section as compared to $qg\to\gamma q$, and b) the initial-state interactions contribute a further suppression by a factor of eight with respect to the standard partonic cross-section.

As was the case with the GPM estimates, the results for the inclusive photon asymmetry are somewhat lower compared to the direct photon asymmetry.

\begin{figure}[t]
\vspace*{-1cm}
\begin{center}
\includegraphics[width=\linewidth]{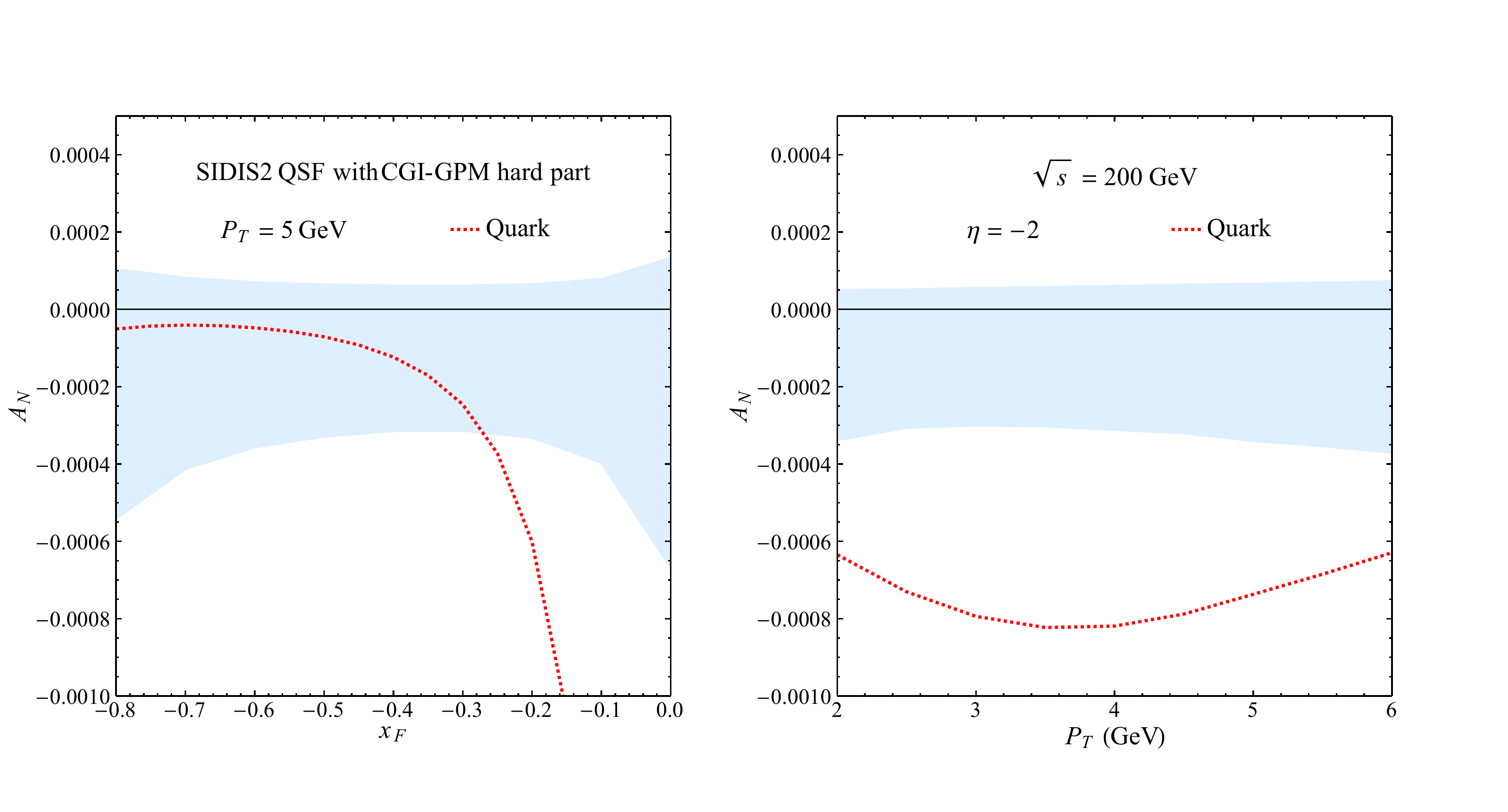}
\vspace*{-1.5cm}
\caption{Burkardt Sum Rule based constraints on $A_N$ in the CGI-GPM framework and predictions of asymmetry using the SIDIS2~\cite{Anselmino:2008sga} QSFs. The light blue band shows the envelope of asymmetries obtained from GSFs that obey the BSR based constaint, Eq.~23.}
\label{CGIGPMANband}
\end{center}
\end{figure}

\subsubsection{Asymmetry estimates using existing fits as well as constraints from the Burkardt Sum Rule}

In Fig.~\ref{CGIGPMANband} we plot the range of values for the GSF contribution to the asymmetry in the CGI-GPM framework, as allowed by the constraint from the Burkardt Sum Rule. We also show the quark contribution to the asymmetry as given by the SIDIS2 QSFs. Since there are no available fits of the GSF in the CGI-GPM framework, we do not show any gluon contribution, except for the asymmetry band allowed by the BSR constraint. It is important to note that the band shown in the plot corresponds only to the contribution of the $f$-type GSF. As mentioned in Sec.~III, the BSR does not constrain the $d$-type Sivers function as it is odd under charge conjugation.  As was the case with the GPM framework, we find that both the fits as well as the constraints based on the Burkardt Sum Rule predict negligible values for the gluon contribution to the asymmetry.

Overall, in the CGI-GPM framework we see that asymmetry predictions decrease in magnitude but the relative importance of the gluon contribution increases in the highly backward regions, i.e., $x_F<-0.3$.

\section{Conclusions}

In this work we have presented results for SSA in the production of prompt photons in the negative rapidity region at RHIC.  In this region, the production of direct photons is dominated by gluons in the transversely polarised proton through the QCD Compton process, $gq\to\gamma q$. Thus any observed asymmetry in this region could be a strong indication of a non-zero gluon Sivers function. We find that the use of a gluon Sivers function that saturates the positivity bound can lead to an asymmetry in direct photons of upto 10\%. We find that inclusion of fragmentation photons can dilute the asymmetries by anywhere between 10-50\%  of the value for just the direct photons. A stricter constraint on the GSF than the positivity bound can be obtained using the Burkardt Sum Rule. In Ref.~\cite{Anselmino:2008sga}, wherein the SIDIS2 QSFs were extracted, it was found that the BSR allowed an average gluon transverse momentum in the range $-10\leq\langle k_{\perp g}\rangle\leq48\text{ MeV}$. We find that the GSF parametrisations that satisfy this constraint give negligible asymmetries.

Further we also studied the asymmetry in the context of the Colour-Gauge Invariant Generalised Parton Model, in which the non-universality of the Sivers functions are accounted for by taking into account the effects of the process dependent initial state and final state interactions. We find that both the f-type and d-type GSFs  can contribute to an SSA in direct photon production. Both of them, when saturated, lead to peak direct photon asymmetries of around 5\%.   Constraints based on the BSR apply only to the $f$-type GSF and give negligible values for the asymmetry, as was the case with the GPM framework. However $d$-type GSF is so far not constrained by anything except the positivity bound and hence, in principle, can be much larger.

\section{Acknowledgements}

The work of R.M.G. is supported by the Department of Science and
Technology, India under Grant No. SR/S2/JCB-64/2007 under the J.C. Bose Fellowship scheme.  A.M would like to thank the Department of Science and Technology, India for financial support under Grant No.EMR/2014/0000486. RMG wishes to acknowledge the support of the Institute of Physics and Astronomy, Amsterdam for the IPA Visiting Professorship in 2018 and the hospitality of the theory group at NIKHEF, Amsterdam.

\section{Appendix}
In this work, we have considered the single-inclusive production of both direct photons as well as photons produced via fragmentation of partons. Here we oultine the treatment of parton kinematics for both cases.

The momenta of the polarised proton (A), unpolarised proton (B) and the photon can be written in the $pp$ centre of mass frame as,
\be
p_A=\frac{\sqrt{s}}{2}(1,0,0,1),~p_B=\frac{\sqrt{s}}{2}(1,0,0,-1)~\text{and}~p_\gamma=(E_\gamma=\sqrt{P_T^2+P_L^2},P_T,0,P_L)
\ee

The parton from the polarised proton (a) and the parton from the unpolarised proton (b) carry light-cone momentum fractions $x_a=p_a^+/p_A^+$ and $x_b=p_b^-/p_B^-$ and transverse momentua $\bfk_{\perp a}$ and $\bfk_{\perp b}$ respectively. Taking them both to be on-shell, their momenta are given by,
\bea
p_a&=x_a\frac{\sqrt{s}}{2}\left(1+\frac{k_{\perp a}^2}{x_a^2 s},\frac{2k_{\perp a}}{x_a\sqrt{s}}\cos\phi_{\perp a},\frac{2k_{\perp a}}{x_a\sqrt{s}}\sin\phi_{\perp a},1-\frac{k_{\perp a}^2}{x_a^2 s}\right)
\\\nonumber
p_b&=x_b\frac{\sqrt{s}}{2}\left(1+\frac{k_{\perp b}^2}{x_b^2 s},\frac{2k_{\perp b}}{x_b\sqrt{s}}\cos\phi_{\perp b},\frac{2k_{\perp b}}{x_b\sqrt{s}}\sin\phi_{\perp b},-1+\frac{k_{\perp b}^2}{x_b^2 s}\right)
\label{incomingparton}
\eea
where $\phi_{\perp a}$ and $\phi_{\perp b}$ are the azimuthal angles of the parton transverse momenta.

In case of direct photon production, the treatment of parton kinematics is relatively simple as the on-shell condition $\shat+\that+\uhat=0$ can be used to fix one of the variables, such as $x_a$ or $x_b$. In addition, we have the requirement that the energy of the incoming parton should not be greater than that of its parent particle, $E_{a(b)}\leq E_{A(B)}$. This leads to the following constraint,
\be
k_{\perp a(b)}<\sqrt{s}~\min[x_{a(b)}, \sqrt{x_{a(b)}(1-x_{a(b)})}].
\ee

In case of photon production via fragmentation, the transverse momentum in the fragmentation makes the kinematics more involved. In this case, the photon is produced via fragmentation of the final state parton in a 2-to-2 process, $ab\to cd$. The momentum of the photon, relative to the fragmenting parton is given by $z$, 
the light-cone momentum fraction of the heavy meson and $\bfk_\gamma$, the transverse momentum of the meson with respect to direction of heavy quark. In a choice of coordinates where the fragmenting parton's momentum, $p_c$ is along the $z$-axis, the photon's momentum can be written as
\be
p_\gamma=(E_\gamma,0,0,|\bfp_\gamma - \bfk_\gamma|) +  (0,\bfk_\gamma)
\ee
where the first term on the right is the component along the fragmenting parton's direction and the second term is the component transverse to it. Here, $\bfk_\gamma$ is simply $(k_{\gamma_x},k_{\gamma_y},0)=(\bfk_{\gamma_\perp},0)$. In the lab coordinates however, $\bfk_\gamma$ can have all three components non-zero and is specified as,
\be
\bfk_\gamma=k_\gamma(\sin\theta\cos\phi,\sin\theta\sin\phi,\cos\theta) \text{, with } |\bfk_\gamma|=|\bfk_{\gamma_\perp}|
\ee
and the orthogonality condition $\bfk_\gamma.\bfp_c=0$ ensures that $\bfk_\gamma$ lies in a plane perpendicular to $\bfp_c$. The light-cone momentum fraction $z$ is given by,
\be
z=\frac{p_\gamma^+}{p_c^+}=\frac{E_\gamma+|\bfp_\gamma-\bfk_\gamma|}{E_c+|\bfp_c|}=\frac{E_\gamma+\sqrt{\bfp_\gamma^2-\bfk_\gamma^2}}{2E_c}
\label{light-cone-z}
\ee
This gives us the expression for the energy of the heavy quark,
\be
E_c=\frac{E_\gamma+\sqrt{\bfp_\gamma^2-\bfk_\gamma^2}}{2z}.
\label{energy-relation}
\ee
The expression for $\bfp_c$ can be obtained from the fact that it is collinear with $\bfp_\gamma - \bfk_\gamma$ and that the unit vector constructed out of both must therefore be equal,
\be
\bfp_c=\frac{E_c}{E_\gamma-k_\gamma^2}\left(P_T-k_\gamma\sin\theta\cos\phi,-k_\gamma\sin\theta\sin\phi,P_L-k_\gamma\cos\theta\right)
\label{momentumrelation}
\ee

where we have used the orthogonality condition, $\bfk_\gamma.\bfp_c=0$. Eqs. \ref{energy-relation} and \ref{momentumrelation} relate the energy and momentum of the observed photon with that of the fragmenting parton for given values of $k_\gamma$ and $z$. 

Using the expressions for the parton momenta given in Eqs.~\ref{incomingparton} and \ref{momentumrelation}, one can solve the on-shell condition $\shat+\that+\uhat=0$ for $z$~\cite{DAlesio:2004eso}.

The term $d^3 \bfk_\gamma \, 
\delta (\bfk_\gamma \cdot \hat{\bfp}_c)$ in Eqs. \ref{fragden} and \ref{fragnum} ensures that the $\bfk_\gamma$ integration is only over momenta transverse to the fragmenting parton:
\be
d^2\bfk_{\gamma_\perp}=d^3 \bfk_\gamma \, 
\delta (\bfk_\gamma \cdot \hat{\bfp}_c)=dk_\gamma\text{ }k_\gamma\text{ }d\theta \text{ }d\phi\frac{|\bfp_\gamma-\bfk_\gamma|}{P_T\sin\phi_1}\left[\delta(\phi-\phi_1)+\delta(\phi-(2\pi-\phi_1))\right]
\label{eeeee}
\ee
where,
\be
\cos \phi_1=\frac{k_\gamma-P_L\cos\theta}{P_T\sin\theta}
\ee
Limits on $k_\gamma$ can be obtained by requiring $|\cos \phi_1|\leq1$,
\be
\text{max}\left[P_L\cos\theta-P_T\sin\theta,0\right]\leq k_\gamma\leq\text{max}\left[P_L\cos\theta+P_T\sin\theta,0\right].
\ee

\end{document}